\documentclass[aps,pre,twocolumn,superscriptaddress,showpacs,showkeys ]{revtex4-1}

\usepackage{graphicx}                   
\usepackage{hyperref}                   
\usepackage{amsmath,amssymb}            
\usepackage{epstopdf}
\usepackage{subcaption}					

\usepackage{color}
\definecolor{orange}{rgb}{1,0.5,0}
\definecolor{brown}{rgb}{0.65, 0.16, 0.16}
\definecolor{phlox}{rgb}{0.87, 0.0, 1.0}

\graphicspath{{figs/}}					

\begin{document}

\title{On the role of anaxonic local neurons in the crossover to continuously varying exponents for avalanche activity}

\author{M. Rahimi-Majd}
\affiliation{Department of Physics, Shahid Beheshti University, 1983969411, Tehran, Iran}
\email{rahimimajd.milad@gmail.com}

\author{M. A. Seifi}
\affiliation{Department of Physics, University of Mohaghegh Ardabili, P.O. Box 179, Ardabil, Iran}
\email{adelseify86@gmail.com}

\author{L. de Arcangelis}
\affiliation{Dept. of Engineering,
	University of Campania Luigi Vanvitelli, 81031 Aversa (CE), Italy}
\email{lucilla.dearcangelis@unicampania.it }

\author{M. N. Najafi}
\affiliation{Department of Physics, University of Mohaghegh Ardabili, P.O. Box 179, Ardabil, Iran}
\email{morteza.nattagh@gmail.com}

\begin{abstract}
Local anaxonic neurons with graded potential release are important ingredients of nervous systems, present in the olfactory bulb system of mammalians, in the human visual system, as well as in arthropods and nematodes. We develop a neuronal network model including both axonic and anaxonic neurons and monitor the activity tuned by the following parameters: The decay length of the graded potential in local neurons, the fraction of local neurons, the largest eigenvalue of the adjacency matrix and the range of connections of the local neurons. Tuning the fraction of local neurons, we derive the phase diagram including two transition lines: A critical line separating subcritical and supercritical regions, characterized by power law distributions of avalanche sizes and durations, and a bifurcation line. We find that the overall behavior of the system is controlled by a parameter
tuning the relevance of local neuron transmission with respect to the axonal one. The statistical properties of spontaneous activity are affected by local neurons at large fractions and in the condition that the graded potential transmission dominates the axonal one. In this case the scaling properties of spontaneous activity exhibit continuously varying exponents, rather than the mean field branching model universality class.
\end{abstract}


\maketitle

\section{Introduction}
Neurons vary in shape and size and can be classified on the basis of their morphology and function. In particular, spiking neurons are cells that elicit an action potential along the axon, according to
an all-or-none behavior depending on their membrane potential. However, also neurons without axon exist. These neurons, classified as anaxonic, have only dendrites and can communicate only with their closest neighbors, for which reason they are often called “local neurons” \cite{kalat2015biological}. Since they cannot produce an action potential, these neurons do not have an all-or-none behavior but, once stimulated, they exhibit a graded potential, a membrane potential that varies in magnitude proportionally to the intensity of the stimulus. This graded potential leads to a continuous neurotransmitter release at dendro-dendritic synapses, conveying information to nearby neurons in all directions and spatially decaying in amplitude with the distance. The continuous neurotransmitter release necessarily requires a continuous generation and reuptake of vesicles to meet the high demand rate. This kind of neurons is very complex to study since, because of their small size, 
it is very difficult to insert an electrode inside. As a consequence, it is still not clear the value of their percentage in different neuronal systems.
Although few of the molecular mechanisms acting at graded synapses are known, the fundamental processes characterizing spiking and graded synapses seem to be similar~\cite{juu}.\\

Nematodes are an example of animals which have non-spiking neurons which transmit information only through electrical and synaptic graded transmission~\cite{liu2018c}. In arthropods, each segmental ganglion is formed by a small number of cells, where local neurons play an important role~\cite{smarandache2016arthropod}. In the olfactory bulb of mammalians like mice, the Granule cells, as a gabaergic type ones, are anaxonic and constitute the largest population of interneurons in this part of the central nervous system~\cite{licausi2018role}. In the human visual system, retinal rods cells lack axons and produce graded potentials~\cite{bhandawat2010signaling}. Similar structures have been observed in the bumblebee visual system~\cite{rusanen2017characterization}  and in the vertebrate and invertebrate retina~\cite{juu}. Local, non-spiking neurons have also been found to play a central role in the motor system of crustacea~\cite{mendel}, crabs~\cite{dicaprio}, insects~\cite{burr}, where spiking and non-spiking neurons are found to operate in synergy to modulate the response to stimuli. Because of their continuous changes in membrane potential, non-spiking neurons are found to have a higher information transfer rate than spike-mediated transmission, namely, having a higher signal to noise ratio, they are able to transfer more information over short time intervals ($\sim 100$ms) than spiking neurons~\cite{dicaprio}. Moreover, they might have an important role not only in response modulation but also in memory and learning because of their higher fidelity in encoding information.\\

Spontaneous brain activity, related to electrophysiological processes taking place in the absence of specific tasks and external stimuli, has recently
revealed a complex bursty behavior observed at different scales and in different neuronal systems, from dissociated neurons to the human brain ~\cite{beggsPlenz2003, mazzoni2007, pasquale2008, gireesh2008, petermann2009, haimovici2013, shriki2013, critrev}. 
Bursts in activity have been named neuronal avalanches and their statistics has been widely investigated in both experimental and numerical datasets. Interestingly, a very robust scaling behavior is observed for the distributions of the avalanche sizes $S$ and durations $D$, according to $P(z)\propto z^{-\tau_z}$, where $z=S,D$ and $\tau_z$ is the corresponding exponent. These assume consistently the values $\tau_S\simeq 1.5$ and $\tau_D\simeq 2.0$ characterizing the mean field branching process universality class \cite{zapperi1995}. However, also different exponent values have been measured in experiments, like M/EEG recordings ~\cite{shriki2013neuronal,dalla2019modeling,palva2013neuronal,zhigalov2015relationship,yaghoubi2018neuronal}. The role of local neurons in the scaling behavior of spontaneous activity is still an open question, difficult to address experimentally.\\

\begin{figure}
	\centerline{\includegraphics[scale=0.6]{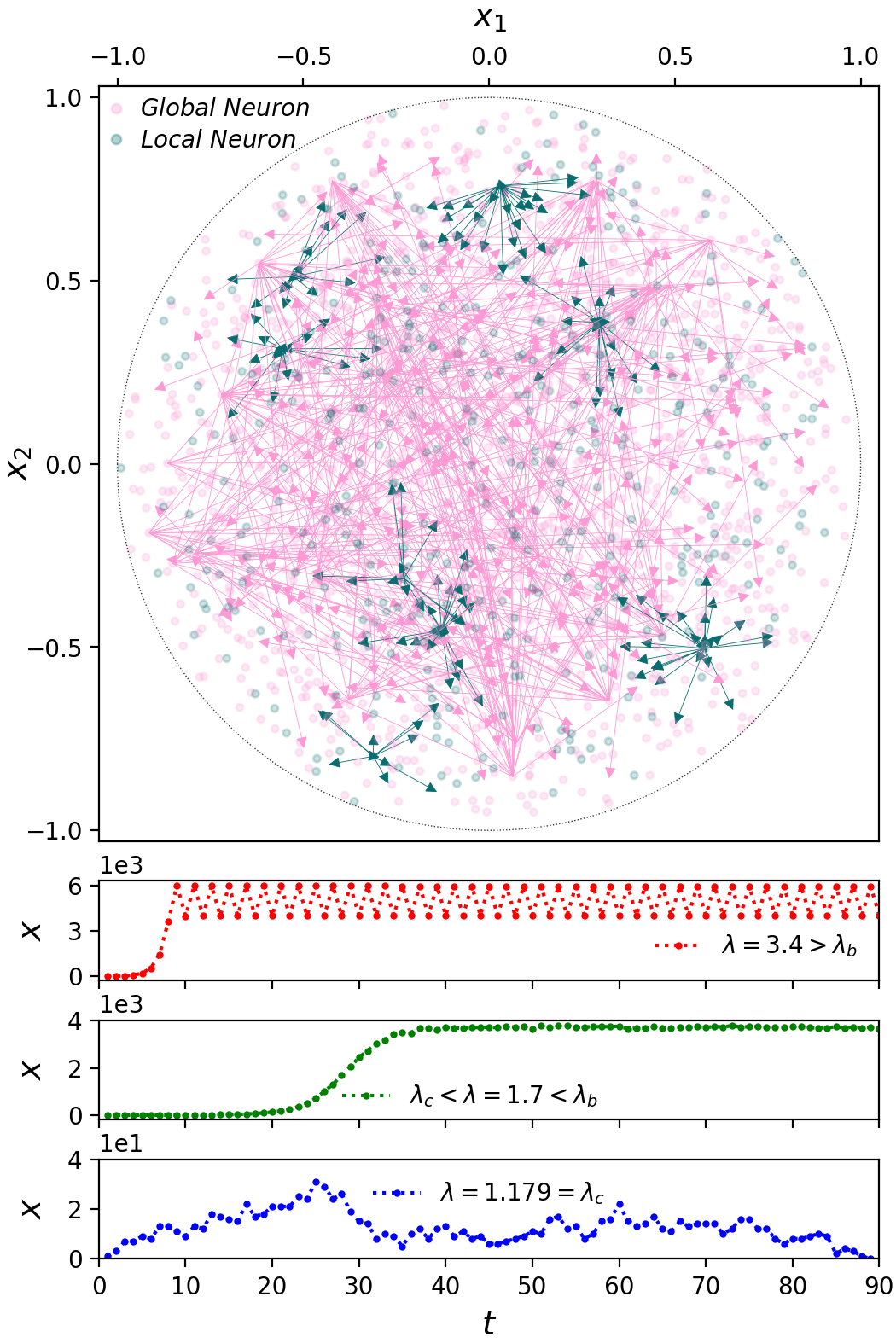}}
	\caption{ Upper panel: A 2D embedding of the network in our model including $2000$ nodes, $ 20\% $ of which are local nodes, with output links of local (green, not exceeding the range $r_1=\frac{R}{3}$) and global (pink) nodes for $q_\zeta=0.02$. Lower panels from tom to bottom: Temporal signal of activity for the oscillatory, supercritical and critical regimes for $\zeta=0.3$ and $N_g=10000$.}
	\label{fig:network}
\end{figure}

The majorities of studies of neuronal models focuses on networks of integrate and fire spiking neurons. Numerical and theoretical approaches have investigated both the spontaneous and evoked activity of such networks, implementing a variety of physiological ingredients, such as short and long-term plastic adaptation, or a complex network structure~\cite{lucilla2006,levina,kanders,dragon,lombchaos,lucillaActD,van2018,lucillaLearning,vittorioLearning,fdt}. Conversely, the investigation of the functional role of non-spiking neurons in the activity of a complex network is a problem which has received little attention in the literature. Moreover, a model with a continuous transition to a single absorbing state, where all neurons are off, and no further symmetries falls into the mean field branching model universality class~\cite{janssen1981nonequilibrium,grassberger1981phase}, also seen for the Linkenkaer-Hansen critical oscillations (CROC) model at the transition line~\cite{poil2012critical}. The important point is that exponents governing avalanche distributions in principle depend on various parameters (such as local threshold defined to identify the start and end of an avalanche) and often show high variability~\cite{shriki2013neuronal,dalla2019modeling,palva2013neuronal,zhigalov2015relationship,yaghoubi2018neuronal}. Recently it was shown that the CROC model shows varying exponents along the critical line~\cite{dalla2019modeling}. Therefore exploring the full variability of the exponents observed experimentally represents a challenge to any model. 

Experimental evidence suggests that the cooperative contribution of spiking and non-spiking neurons can optimize information transmission in several neuronal systems and therefore understanding the role of non-spiking neurons in neuronal activity is an intriguing open question.
Furthermore, the dynamical aspects of graded potential transmission has been modeled~\cite{zetterberg1978performance,kretzberg2001neural}, however its role in the activity of a neuronal culture has not been analyzed yet.
 Here we study the spontaneous activity of a network made of both spiking and non-spiking neurons and, by tuning a number of parameters, as the fraction of the non-spiking neurons, the decay range of the graded potential and the network connectivity, we monitor the different phases of activity unveiling the fundamental contribution provided by anaxonic neurons. Our approach allows to observe a crossover from the mean field branching process universality class to varying-exponents along the critical line, opening a way to comparison with experimental results.\\


\begin{figure}
	\centerline{\includegraphics[scale=0.6]{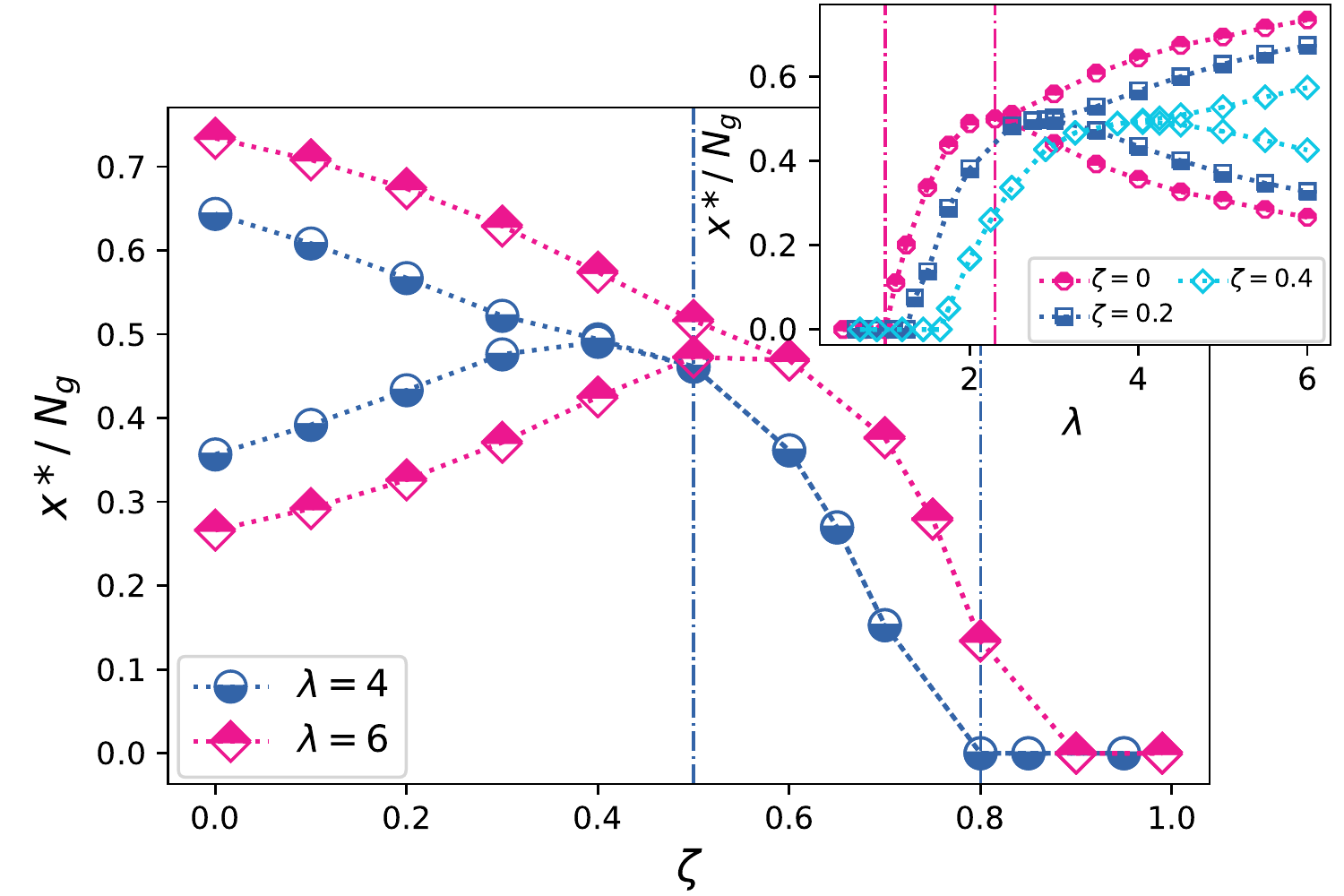}}
	\caption{The fixed points vs. $\zeta$ for $\lambda=4$ and $\lambda=6$. The vertical dashed lines show the critical and bifurcation points for $\lambda=4$. Inset: The same data vs. $\lambda$ for different $\zeta$, i.e. $\zeta=0, 0.2$ and $0.4$. The vertical dashed lines show the critical and bifurcation points for $\zeta=0$. For all curves $N_g=10000$, $r_0=\frac{r_1}{10}$, $r_1=\frac{R}{3}$.}
	\label{fig:fixed_points}
\end{figure}

\section{Model}
Our model consists of two types of excitable units, global and local neurons. The global neurons at each instant of time can be either ``on'' (active) or ``off'' (inactive), whereas the local neurons release graded potentials. Given $N_l$ the number of local neurons (LoN) and $N_g$ the number of global neurons (GlN), the population is controlled by the fraction of LoNs $\zeta\equiv\frac{N_l}{N_t}$ where $N_t\equiv N_l+N_g$ is the total number of neurons. The connection weights are randomly distributed $w_{i,j}\in ]0,2\sigma]$ where $\sigma$ tunes the heterogeneity in synaptic strengths. We recall that for a random network with only $N$ spiking neurons ($\zeta=0$), it was shown~\cite{moosavi2017refractory,restrepo2007approximating} that the dynamics is controlled by the largest eigenvalue of the adjacency matrix of network $\lambda =\sigma \left\langle k \right\rangle$, where $\left\langle k\right\rangle =qN$ is the average node degree, and $q$ is the probability that a connection is established between two random nodes.
In our model we generalize this result to $\lambda_{\zeta}=\sigma q_{\zeta}N_t$, where $q_{\zeta}$ is the probability to establish a connection between \textit{any} pair of nodes. In our model avalanches are defined only through the GlNs, i.e. the instantaneous activity $x(t)$ is defined as the number of firing GlNs at each time $t$. On this basis, in our simulations $N_g$ is the system size, kept fixed, whereas $N_t$ and $N_l$ depend on $\zeta$, i.e. $N_t=\frac{N_g}{1-\zeta}$ and $N_l=\frac{\zeta}{1-\zeta} N_g$ ($\zeta<1$). Therefore, for $\zeta=0$ we have $\lambda_{\zeta=0}=\sigma q_{\zeta=0}N_g$. In order to determine $q_{\zeta}$, we impose that the connectivity of the network in the absence of local neurons ($\zeta=0$) is the same as the connectivity of the network for a given $\zeta$, i.e. we set $\lambda_{\zeta}=\lambda_{\zeta=0}$, leading to $q_{\zeta}=q_{\zeta=0}\frac{N_g}{N_t}$. Noticing that the probability that a generic node is chosen for establishing a connection with any another node (global or local) is $q_{\zeta}(N_t-1)\approx q_{\zeta}N_t$, we conclude that the above condition assures us that this probability does not change by varying $\zeta$. Under these assumptions, we can examine the effect of the population of LoNs without changing the whole connectivity of the network. 
By defining $n_{\zeta}$ as the number of links between any pair of neurons for a fraction $\zeta$ of LoNs, we obtain

	\begin{equation}
	n_{\zeta}=\frac{1}{2}q_{\zeta}N_t(N_t-1)=\frac{1}{2}q_{\zeta=0}N_gN_t.
	\end{equation}
We can then construct the network for a given value of $q_{\zeta=0}$ by randomly distributing directional connections between neurons randomly placed in a sphere  of radius $R$, with the condition for LoNs that the length of their \textit{outgoing} connections cannot be larger than $r_1$, which bounds the range on local connections, see Fig.~\ref{fig:network}. 

The dynamics of LoNs and GlNs are governed by two different equations. The probability that a global neuron spikes at time $t$ is given by~\cite{moosavi2017refractory}
\begin{equation}
\begin{split}
p(A_i(t)=1)=&\delta_{A_i(t-1),0}F\left[\sum_{j'}w_{ij'}A_{j'}(t-1)\right. \\
&\left. +\sum_{j''}w_{ij''}V_{j''}(t-1)e^{-\frac{\left|r_i-r_{j''} \right|}{r_0}}\right]
\end{split}
\label{Eq:firing}
\end{equation}
where $r_i$ and $V_i$ are the position and membrane potential of neuron $i$ and $A_i(t)=1(0)$ characterizes the firing (non firing) state of neuron $i$. 
The first (second) sum is over the global (local) neurons with $F(y)=y$ if $0\leq y\leq 1$ and $F(y)=1$ otherwise.
After firing the potential of global neurons is set to zero and neurons remain in a refractory state for one time step. Conversely, the potential of the LoNs evolves in time according to
\begin{equation}
\begin{split}
V_i(t)&=\sum_{j'}w_{ij'}A_{j'}(t-1)\\
&+\sum_{j''}w_{ij''}V_{j''}(t-1)\exp\left[-\frac{\left|r_i-r_{j''}\right|}{r_0}\right]. 
\end{split}
\label {Eq:sub-threshold}
\end{equation}
Here the first term integrates the membrane potential variations of LoNs due to the firings of connected GlNs, whereas the second term represents the contribution
of the graded potentials released in the previous time step by the neighboring LoNs. Note that GlNs have a refractory time, whereas the LoNs do not, as required by their non-spiking feature.
To avoid the occurrence of very high values of $V_i(t)$, we define a threshold $V_{\text{th}}$ to set a cap on $V_i(t)$. From the physiological point of view, $V_{\text{th}}$ implements that a neural cell can support a maximal potential difference, whose specific value does not affect our final results. We therefore set $V_{\text{th}}=1$. In Eq.s~\ref{Eq:firing} and \ref{Eq:sub-threshold} the decay factor $\exp\left[-\left|r_i-r_{j''}\right|/r_0\right]$, with the distance $\left|r_i-r_{j''}\right|$  between two neurons and $r_0$ a characteristic range, expresses the decay of the graded potential released by a LoN. In simulations we set $q_{\zeta=0}=0.02$, and generate about $5\times 10^5$ independent samples for all sets of parameters, $\lambda$, $\zeta$, $r_0$, $r_1$ and $N_g$, where $\lambda$ is changed by fixing $\sigma$. For each network, we start from a configuration where all nodes are off, and turn on a random node, monitoring the activity, $x(t)$, defined as the number 
of firing global neurons, over several thousand time steps. Moreover,in order to determine the avalanche statistics, we evaluate the avalanche size and duration as the number of firing global neurons in a burst and its temporal extension, respectively. 

\begin{figure*}
	\centering
	\begin{subfigure}{0.49\textwidth}\includegraphics[width=\textwidth]{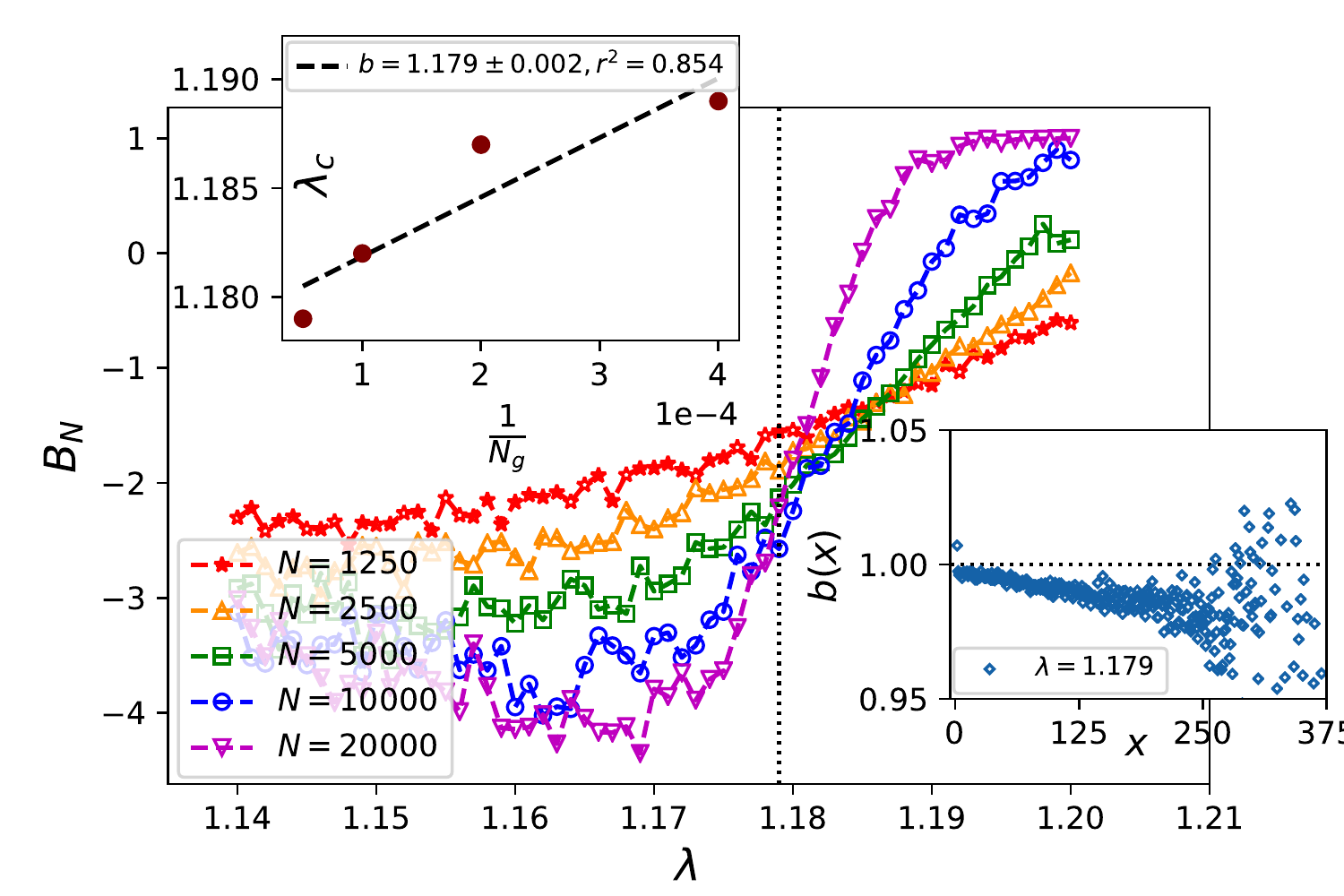}
		\caption{}
		\label{fig:Critical}
	\end{subfigure}
	\begin{subfigure}{0.49\textwidth}\includegraphics[width=\textwidth]{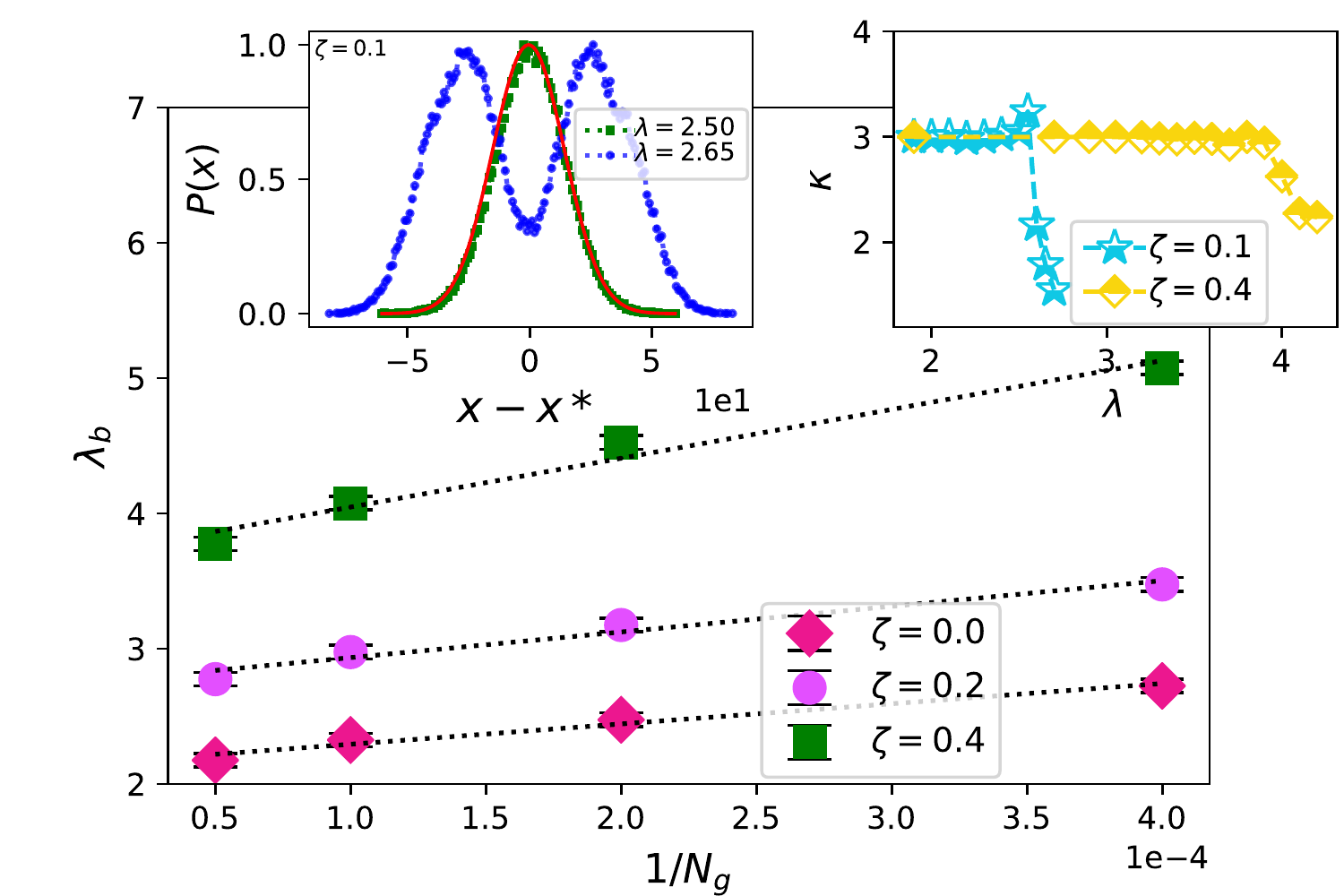}
		\caption{}
		\label{fig:Bifurcations}
	\end{subfigure}
	\caption{(a) Binder cumulant vs. $\lambda$ for various values of $N_g$. Upper inset: The extrapolation of $\lambda_c(N)$ (defined in the text) in the limit $N_g\to \infty$ provides the value $1.179$, shown by the vertical dashed line in main panel. Lower inset: The activity-dependent branching ratio $b(x)$ vs. $x$ for $\lambda=1.179$ and $N_g=20000$. For all data $\zeta=0.3$, $r_0=r_1$ and $r_1=\frac{R}{3}$. (b) The extrapolation of $\lambda_b$ i.e. the bifurcation point for $r_0=\frac{r_1}{10}$ with linear fits (dashed lines). Left inset: distribution function of the instantaneous activity $x(t)$ for $\lambda=2.50$ and $\lambda=2.65$ where the bifurcation point is $\lambda_b=2.55$. The red line is Gaussian fit of $\lambda=2.50$. Right inset: The kurtosis in terms of $\lambda$ for $\zeta=0.1$ and $\zeta=0.4$.}
	\label{fig:Points}
\end{figure*}

\section{Phase diagram and avalanche activity}
Once the temporal activity signal is generated, we define $x^*$ as the fixed point of the activity dynamics, numerically determined by taking the average of $x(t)$ in the long-time regime. The system becomes critical at $\lambda_c$~\cite{moosavi2017refractory,larremore2012statistical}, which is defined as the $\lambda$-value above which $x^*$ becomes different than zero for the first time. 
For $\zeta=0$ and without a refractory period~\cite{moosavi2017refractory}, the model undergoes a continuous, absorbing-active (AA) phase transition at $\lambda_c=1$~\cite{moosavi2017refractory}.  For $\lambda<\lambda_c$ the system is inactive, i.e. it requires an external drive to be activated and the activity value $x^*=0$ is the stable attractor of the dynamics. Conversely, for $\lambda>\lambda_c$ the system is active with a fixed point at infinity, or else at $x=N_t$ for finite systems. When the refractory period is included, the system exhibits a bifurcation at $\lambda=\lambda_{\text{bif}}$, identified by an oscillatory behavior and the critical regime is extended as claimed in~\cite{moosavi2017refractory}. In this case, $x^*$ is identified by taking the average for each of the two branches separately.

\begin{figure*}
	\centering
	\begin{subfigure}{0.45\textwidth}\includegraphics[width=\textwidth]{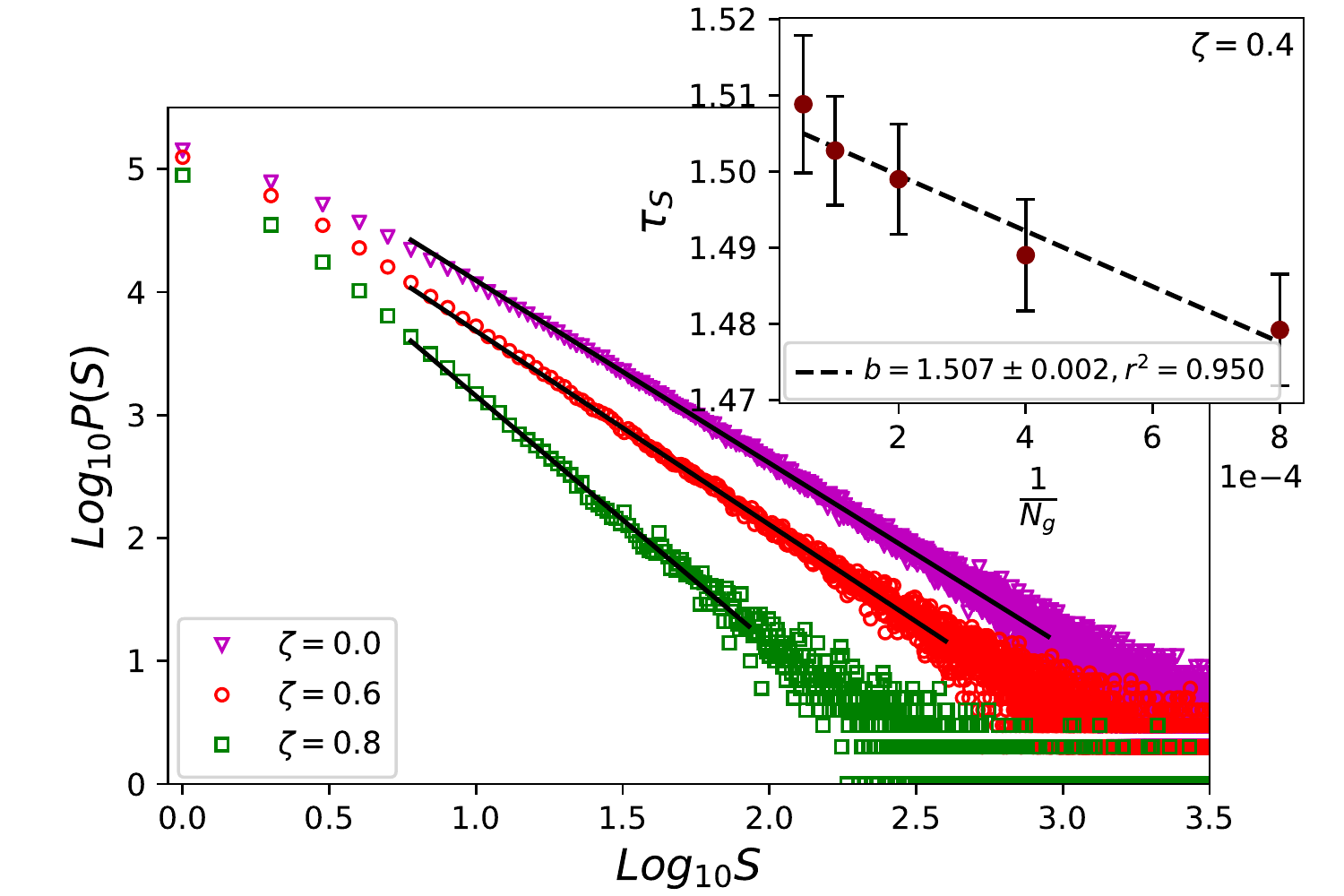}
		\caption{}
		\label{fig:r12}
	\end{subfigure}
	\begin{subfigure}{0.45\textwidth}\includegraphics[width=\textwidth]{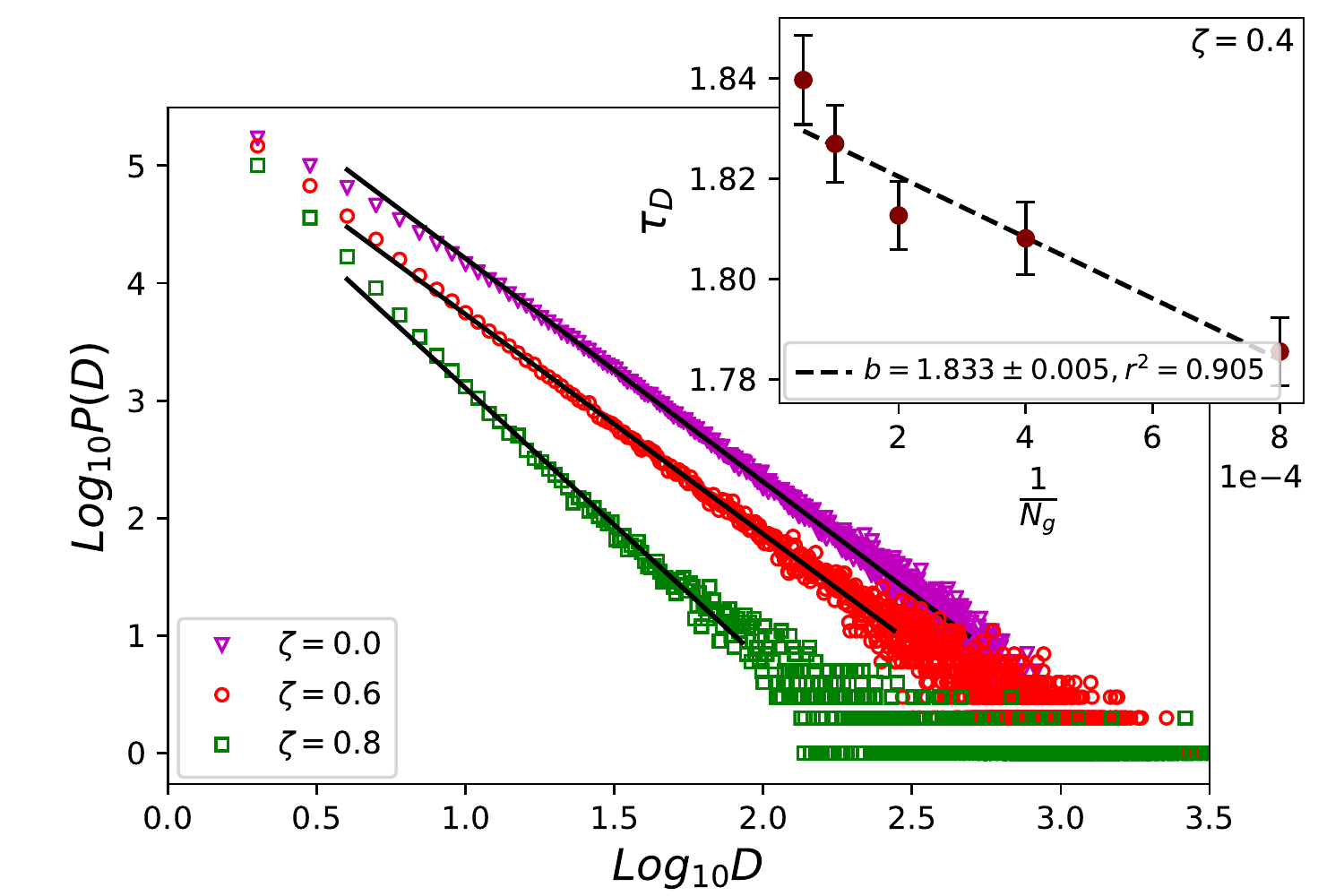}
		\caption{}
		\label{fig:r1div02}
	\end{subfigure}	
	\caption{(a) The distributions of avalanche sizes $S$ (a) and durations $D$ (b) for various fractions of LoNs $\zeta$, $N_g=20000$ and $r_0=r_1=\frac{R}{3}$. Inset: Extrapolation to $N_g\to\infty$ of the corresponding exponents with the linear fit (dashed lines).}
\end{figure*}

Here we find a $\zeta$-driven (LoN-driven) AA continuous phase transition and $\zeta$-driven bifurcation, as evidenced by the three temporal activity signals in the bottom panels of Fig.~\ref{fig:network} corresponding to oscillatory, supercritical and critical behavior. In Fig.~\ref{fig:fixed_points} the bifurcation diagram for two $\lambda$ values indicates that, as $\zeta$ decreases, the system passes three distinct phases:
The sub-critical absorbing state, the super-critical (or extended critical as claimed in~\cite{moosavi2017refractory}) state, and the oscillatory phase. The behavior of the system in terms of $\lambda$ for various $\zeta$-values is shown in the inset: we confirm that for $\zeta=0$ the critical behavior is observed for moderate values of the connectivity level and that an oscillatory behavior sets in as $\lambda$ increases. In order to identify the critical points we use two methods, the Binder cumulant and the branching ratio. The Binder cumulant is defined as
\begin{equation}
B_{N_g}=1-\frac{\left\langle x^4\right\rangle_{N_g} }{3\left\langle x^2\right\rangle_{N_g}^2 }
\end{equation}
which becomes $N_g$-independent at the critical point. In this equation $\left\langle \right\rangle_{N_g}$ is the ensemble average for system size $N_g$. By extrapolating to infinite system size the crossing of successive couples of curves with increasing $N_g$ we obtain the correct value of $\lambda_c$, as shown in the main panel of Fig.~\ref{fig:Critical} and its upper inset. In the lower inset we show the branching ratio for the same parameters, defined by $b(y)\equiv \textbf{E}\left[\frac{x(t+1)}{y}|x(t)=y \right] $ where $\textbf{E}\left[\dots \right] $ is the expectation value. The stable (unstable) fixed point of the dynamics is obtained by imposing $\lim_{y\rightarrow y^*}b(y)=1$ and $\left. \frac{\text{d}b}{\text{d}y}\right|_{y=y^*}<0 $ ($\left. \frac{\text{d}b}{\text{d}y}\right|_{y=y^*}>0 $). One can detect the critical point by inspecting the conditions under which $y^*=0$ \cite{moosavi2017refractory,alstrom1988mean}, i.e. by imposing $\lim_{y\rightarrow 0} b(y)=1$. 

\begin{figure*}
	\centering
	\begin{subfigure}{0.497\textwidth}\includegraphics[width=\textwidth]{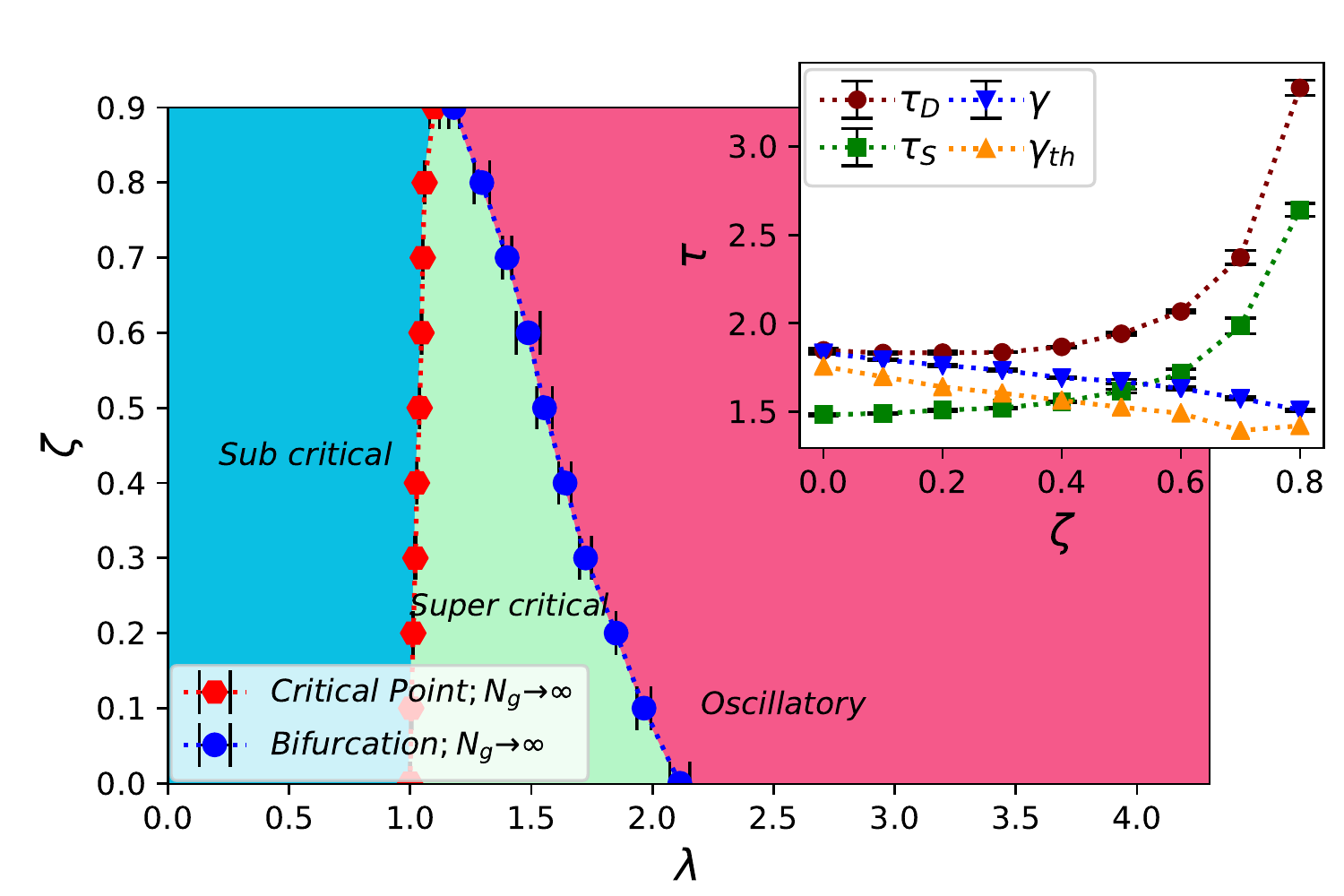}
		\caption{}
		\label{fig:r1mult10}
	\end{subfigure}
	\begin{subfigure}{0.497\textwidth}\includegraphics[width=\textwidth]{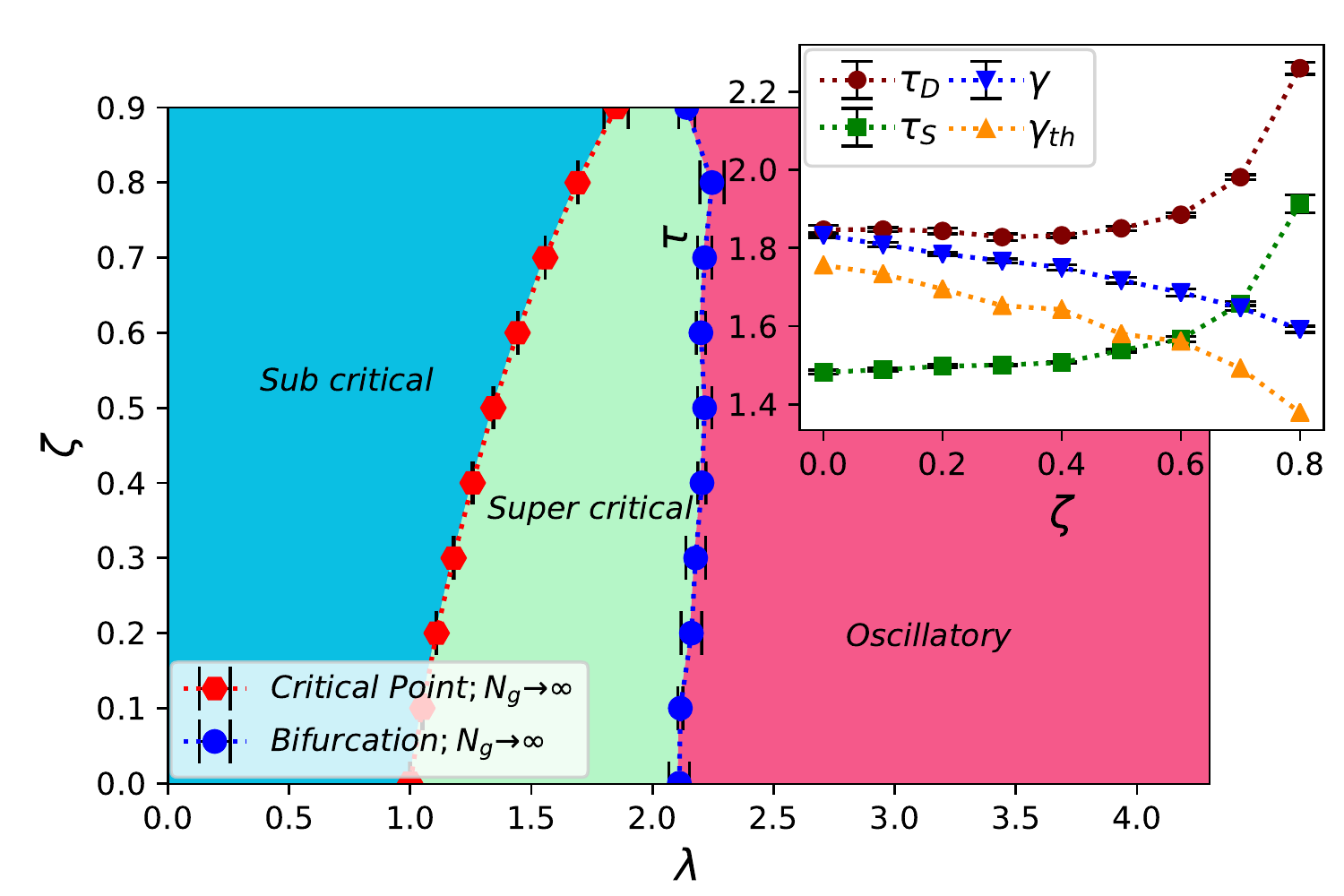}
		\caption{}
		\label{fig:r1}
	\end{subfigure}
	\begin{subfigure}{0.497\textwidth}\includegraphics[width=\textwidth]{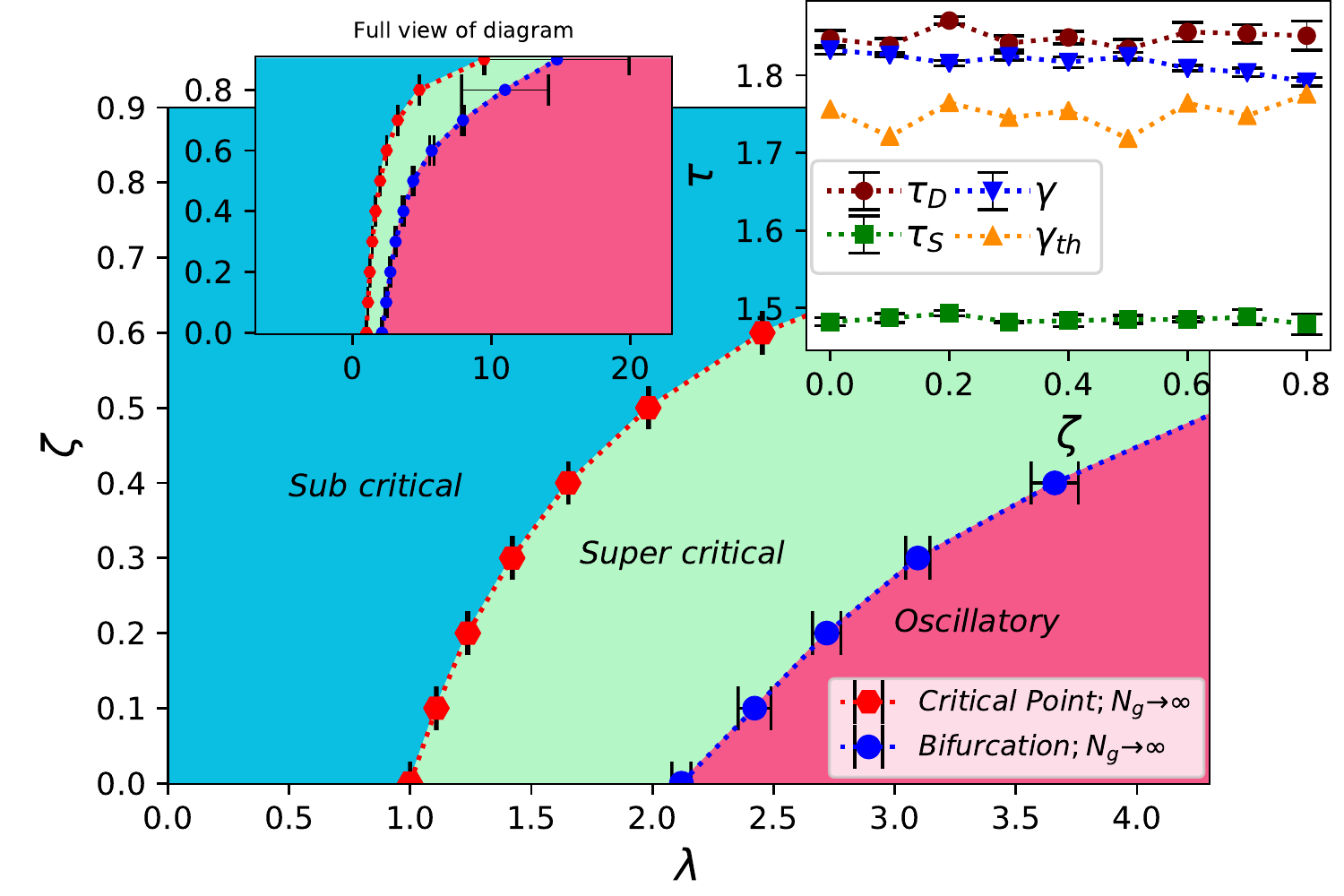}
		\caption{}
		\label{fig:r1div0}
	\end{subfigure}
	\begin{subfigure}{0.497\textwidth}\includegraphics[width=\textwidth]{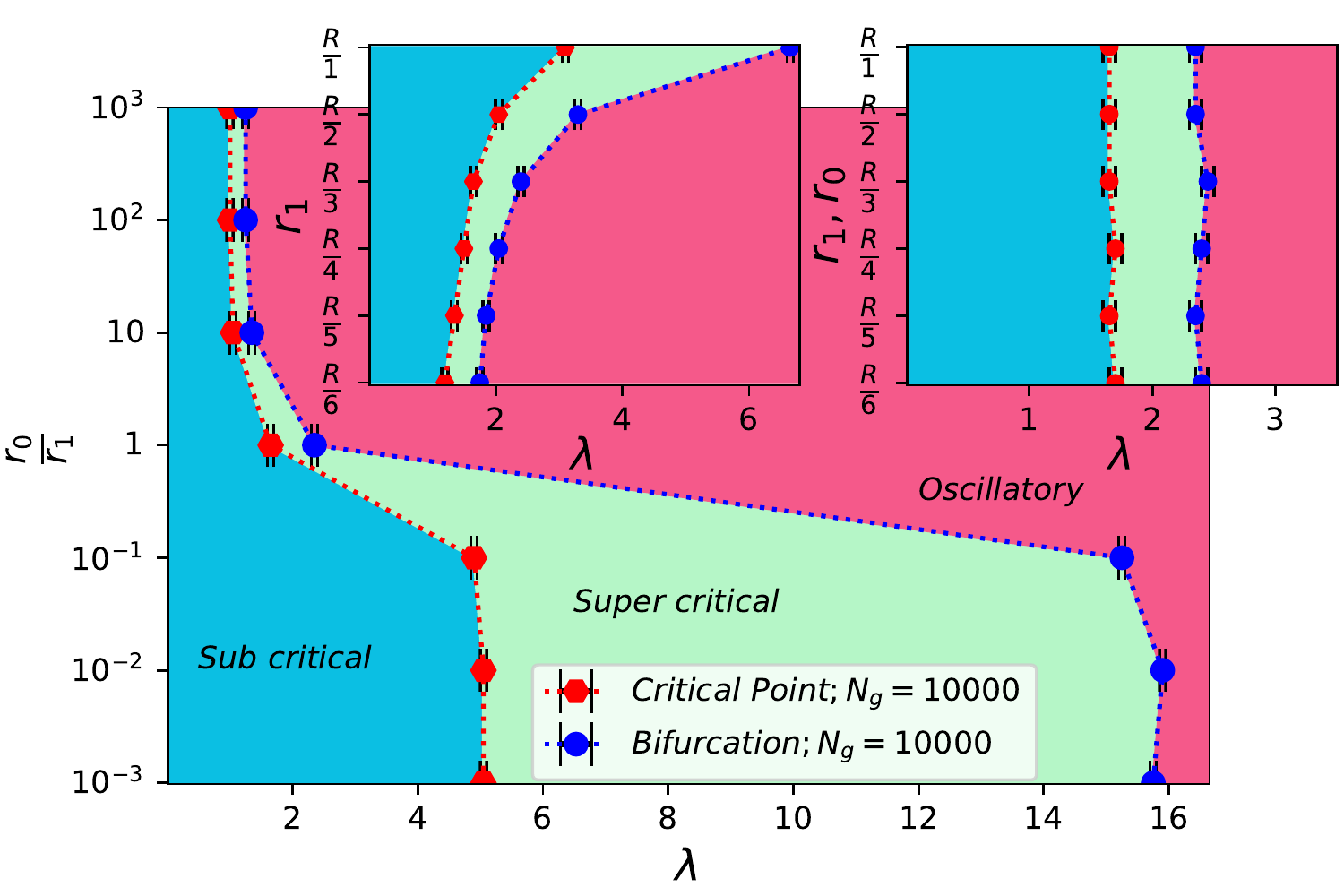}
		\caption{}
		\label{fig:zeta}
	\end{subfigure}
	\caption{(Color online): The two-dimensional ($\zeta-\lambda$) phase diagram in the thermodynamic limit $N_g\rightarrow\infty$ for three $r$ values, i.e. (a) $r=10$, (b) $r=1$, and (c) $r=\frac{1}{10}$. The red circles are critical points and the blue circles are the bifurcation points in the thermodynamic limit. The simulations have been done only on the symbols, and the dashed lines are for eye guide. The insets show the corresponding critical exponents in the thermodynamic limit $N_g\rightarrow\infty$. (d) The two-dimensional $(\lambda,r)$ phase diagram for $N_g=10000$ and $\zeta=0.8$. The insets are the same in terms of $r_1$ for fixed $r_0=\frac{R}{3}$ (left) and for $r_0=r_1$ (right).}
	\label{fig:phases}
\end{figure*}

In order to determine the bifurcation point, we consider that in the supercritical phase the probability distribution of $x$ is a Gaussian (see left inset Fig.\ref{fig:Bifurcations}) and deviates from Gaussianity when the system enters the oscillatory phase from the supercritical phase~\cite{najafi2019effect}. We exploit this result and use the kurtosis to identify the bifurcation points, $\kappa =\sigma_y^{-4}\left\langle y^4 \right\rangle$, where $y=x-\left\langle x \right\rangle$, and $\sigma_y=\sqrt{\left\langle y^2\right\rangle}$. Since $k=3$ for the Gaussian distribution, we define the bifurcation point as the point where the kurtosis deviates from $3$ more than $10\%$ for different system size $N_g$, as shown in the right inset of Fig.~\ref{fig:Bifurcations}. Then by extrapolation, we find the bifurcation points in the thermodynamic limit.
We monitor along the critical line the statistics of avalanches, namely the distribution of sizes and durations. In the Fig.~\ref{fig:r12} and~\ref{fig:r1div02} we show the distributions of the avalanche sizes $S$ and durations $D$, with the corresponding exponents
shown in the upper insets in terms of $N_g$. We also evaluate the scaling exponent $\gamma$ from $S\propto D^\gamma$, whose expected theoretical value for the crackling noise is $\gamma_{th}=\frac{\tau_D-1}{\tau_S-1}\simeq 2$ \cite{sethna}. The values of the exponents will be evaluated along the critical lines.\\

The phase diagrams are shown in Fig.~\ref{fig:phases}, where the transition lines report the values obtained in the thermodynamic ($N_g\rightarrow\infty$) limit. The overall behavior of the system is tuned by the control parameter $r\equiv\frac{r_0}{r_1}$, namely the ratio between the decay range for the graded potential of local neurons and the connectivity range, controlling the influence of LoNs with respect to GlNs. We plot the $(\lambda,\zeta)$ phase diagrams for three cases corresponding to a decreasing relevance of LoNs, $r=10$, $r=1$ and $r=\frac{1}{10}$ in Figs.~\ref{fig:r1mult10},~\ref{fig:r1} and~\ref{fig:r1div0} respectively. For the case $r>r^*\approx 1$ (Fig.~\ref{fig:r1mult10}), the critical line is robust against $\zeta$. The exponents slowly vary for $\zeta\lesssim 0.5$, whereas they show sharper variations for larger $\zeta$, due to the large value of $r$, which lead us to refer to this phase as the varying-exponent phase. This behavior is often seen in experimental results where exponents show high variability~\cite{shriki2013neuronal,dalla2019modeling,palva2013neuronal,zhigalov2015relationship,yaghoubi2018neuronal}.
For the case $r\simeq r^*$ (Fig.~\ref{fig:r1}) the critical line slowly varies with $\zeta $ whereas the bifurcation line is quite stable. The critical exponents also appear to be quite stable for $\zeta\lesssim 0.4$ whereas they show sharp variations for larger fractions of LoNs.
 Finally, for the case $r<r^*$ (Fig.~\ref{fig:r1div0}), the critical and bifurcation lines are sensitive to changes in $\zeta$ and $\lambda$ whereas the exponents are robust, in agreement with the mean-field branching model universality class ($\tau_S=\frac{3}{2}$ and $\tau_D=2$). To monitor more explicitly the behavior in terms of $r$, we plot the $(r,\lambda)$ phase diagram for large $\zeta$ in Fig.~\ref{fig:zeta}, where an abrupt change of behavior is seen at $r^*$: A more extensive super-critical phase is observed for $r<r^*$, namely in the case the decay range of graded potential of LoNs is smaller than the synaptic connectivity range.
Conversely, for $r>r^*$, when the role of LoNs becomes more relevant, a dominant oscillatory phase is detected. The up-right inset shows the diagram for $r=r^*$, revealing that the transition lines are robust with respect to change of $\zeta$. \\

\section{Conclusion}
To summarize, we developed a dynamic neuronal model on a random network which analyses the role played by local neurons on the network activity. Since local neurons do
not follow all or none behavior, their contribution is continuous in time and can affect in a relevant way the spiking activity of global neurons. We tune the relevance of local neurons in terms of their fraction, decay range of the graded potential and connectivity level. Experimental data do not clearly evidence the percentage of local neurons. In some systems, as Granule cells in the olfactory bulb, neurons are anaxonic, forming dendro-dentritic synapses via spiny processes~\cite{licausi2018role}. Moreover they appear to be a robust sub-population in severe diseases: For instance,  investigating the olfactory dysfunction is early stage, non-motor symptom which occurs in 95\% of Parkinson’s disease patients, it was evidenced that the small anaxonic sub-population, continuously replenished by neurogenesis, was moderately reduced in number, much less compared with other neurons in the mid-brain \cite{pass}.For this reason, we explore the phase diagram for a wide range of parameters identifying the different phases. Concerning the scaling behavior of avalanche activity, data confirm that for fractions larger than 50\%, LoNs can indeed affect the values of the critical exponents in the case $r\geq 1$, namely if the decay range of graded potentials is larger or comparable to the synaptic connectivity range. Conversely, for $r<1$ the influence of LoNs becomes more localized and does not sensibly affect the overall network behavior. Moreover, the relevance of the decay range of graded potentials is confirmed by the analysis of the $(r,\lambda)$ phase diagram, suggesting that if
graded potentials decay over an extensive spatial range, LoNs favor synchronization among neurons enhancing the oscillatory phase. The observation of a critical line with varying critical exponents is not a new result in critical phenomena. In the renormalization group approach, the critical point is a fixed point determined by  relevant scaling variables with critical exponents independent of irrelevant variables. The presence of marginal operators makes possible a continuous variation of critical exponents. This result then suggests that LoNs might play the role of marginal variables in the scaling behavior of spontaneous activity.


\noindent

\section{Acknowledgements}	LdA would like to thank MIUR project PRIN2017WZFTZP and the Program VAnviteLli pEr la RicErca: VALERE 2019 for financial support.
 
\bibliography{refs}

\begin{thebibliography}{46}%
\makeatletter
\providecommand \@ifxundefined [1]{%
 \@ifx{#1\undefined}
}%
\providecommand \@ifnum [1]{%
 \ifnum #1\expandafter \@firstoftwo
 \else \expandafter \@secondoftwo
 \fi
}%
\providecommand \@ifx [1]{%
 \ifx #1\expandafter \@firstoftwo
 \else \expandafter \@secondoftwo
 \fi
}%
\providecommand \natexlab [1]{#1}%
\providecommand \enquote  [1]{``#1''}%
\providecommand \bibnamefont  [1]{#1}%
\providecommand \bibfnamefont [1]{#1}%
\providecommand \citenamefont [1]{#1}%
\providecommand \href@noop [0]{\@secondoftwo}%
\providecommand \href [0]{\begingroup \@sanitize@url \@href}%
\providecommand \@href[1]{\@@startlink{#1}\@@href}%
\providecommand \@@href[1]{\endgroup#1\@@endlink}%
\providecommand \@sanitize@url [0]{\catcode `\\12\catcode `\$12\catcode
  `\&12\catcode `\#12\catcode `\^12\catcode `\_12\catcode `\%12\relax}%
\providecommand \@@startlink[1]{}%
\providecommand \@@endlink[0]{}%
\providecommand \url  [0]{\begingroup\@sanitize@url \@url }%
\providecommand \@url [1]{\endgroup\@href {#1}{\urlprefix }}%
\providecommand \urlprefix  [0]{URL }%
\providecommand \Eprint [0]{\href }%
\providecommand \doibase [0]{http://dx.doi.org/}%
\providecommand \selectlanguage [0]{\@gobble}%
\providecommand \bibinfo  [0]{\@secondoftwo}%
\providecommand \bibfield  [0]{\@secondoftwo}%
\providecommand \translation [1]{[#1]}%
\providecommand \BibitemOpen [0]{}%
\providecommand \bibitemStop [0]{}%
\providecommand \bibitemNoStop [0]{.\EOS\space}%
\providecommand \EOS [0]{\spacefactor3000\relax}%
\providecommand \BibitemShut  [1]{\csname bibitem#1\endcsname}%
\let\auto@bib@innerbib\@empty
\bibitem [{\citenamefont {Kalat}(2015)}]{kalat2015biological}%
  \BibitemOpen
  \bibfield  {author} {\bibinfo {author} {\bibfnamefont {J.~W.}\ \bibnamefont
  {Kalat}},\ }\href@noop {} {\emph {\bibinfo {title} {Biological psychology}}}\
  (\bibinfo  {publisher} {Nelson Education},\ \bibinfo {year}
  {2015})\BibitemShut {NoStop}%
\bibitem [{\citenamefont {Juusola}\ \emph {et~al.}(1996)\citenamefont
  {Juusola}, \citenamefont {French}, \citenamefont {Uusitalo},\ and\
  \citenamefont {Weckstr{\"o}m}}]{juu}%
  \BibitemOpen
  \bibfield  {author} {\bibinfo {author} {\bibfnamefont {M.}~\bibnamefont
  {Juusola}}, \bibinfo {author} {\bibfnamefont {A.~S.}\ \bibnamefont {French}},
  \bibinfo {author} {\bibfnamefont {R.~O.}\ \bibnamefont {Uusitalo}}, \ and\
  \bibinfo {author} {\bibfnamefont {M.}~\bibnamefont {Weckstr{\"o}m}},\
  }\href@noop {} {\bibfield  {journal} {\bibinfo  {journal} {Trends in
  neurosciences}\ }\textbf {\bibinfo {volume} {19}},\ \bibinfo {pages} {292}
  (\bibinfo {year} {1996})}\BibitemShut {NoStop}%
\bibitem [{\citenamefont {Liu}\ \emph {et~al.}(2018)\citenamefont {Liu},
  \citenamefont {Kidd}, \citenamefont {Dobosiewicz},\ and\ \citenamefont
  {Bargmann}}]{liu2018c}%
  \BibitemOpen
  \bibfield  {author} {\bibinfo {author} {\bibfnamefont {Q.}~\bibnamefont
  {Liu}}, \bibinfo {author} {\bibfnamefont {P.~B.}\ \bibnamefont {Kidd}},
  \bibinfo {author} {\bibfnamefont {M.}~\bibnamefont {Dobosiewicz}}, \ and\
  \bibinfo {author} {\bibfnamefont {C.~I.}\ \bibnamefont {Bargmann}},\
  }\href@noop {} {\bibfield  {journal} {\bibinfo  {journal} {Cell}\ }\textbf
  {\bibinfo {volume} {175}},\ \bibinfo {pages} {57} (\bibinfo {year}
  {2018})}\BibitemShut {NoStop}%
\bibitem [{\citenamefont
  {Smarandache-Wellmann}(2016)}]{smarandache2016arthropod}%
  \BibitemOpen
  \bibfield  {author} {\bibinfo {author} {\bibfnamefont {C.~R.}\ \bibnamefont
  {Smarandache-Wellmann}},\ }\href@noop {} {\bibfield  {journal} {\bibinfo
  {journal} {Current Biology}\ }\textbf {\bibinfo {volume} {26}},\ \bibinfo
  {pages} {R960} (\bibinfo {year} {2016})}\BibitemShut {NoStop}%
\bibitem [{\citenamefont {LiCausi}\ and\ \citenamefont
  {Hartman}(2018)}]{licausi2018role}%
  \BibitemOpen
  \bibfield  {author} {\bibinfo {author} {\bibfnamefont {F.}~\bibnamefont
  {LiCausi}}\ and\ \bibinfo {author} {\bibfnamefont {N.~W.}\ \bibnamefont
  {Hartman}},\ }\href@noop {} {\bibfield  {journal} {\bibinfo  {journal}
  {International journal of molecular sciences}\ }\textbf {\bibinfo {volume}
  {19}},\ \bibinfo {pages} {1544} (\bibinfo {year} {2018})}\BibitemShut
  {NoStop}%
\bibitem [{\citenamefont {Bhandawat}\ \emph {et~al.}(2010)\citenamefont
  {Bhandawat}, \citenamefont {Reisert},\ and\ \citenamefont
  {Yau}}]{bhandawat2010signaling}%
  \BibitemOpen
  \bibfield  {author} {\bibinfo {author} {\bibfnamefont {V.}~\bibnamefont
  {Bhandawat}}, \bibinfo {author} {\bibfnamefont {J.}~\bibnamefont {Reisert}},
  \ and\ \bibinfo {author} {\bibfnamefont {K.-W.}\ \bibnamefont {Yau}},\
  }\href@noop {} {\bibfield  {journal} {\bibinfo  {journal} {Proceedings of the
  National Academy of Sciences}\ }\textbf {\bibinfo {volume} {107}},\ \bibinfo
  {pages} {18682} (\bibinfo {year} {2010})}\BibitemShut {NoStop}%
\bibitem [{\citenamefont {Rusanen}\ \emph {et~al.}(2017)\citenamefont
  {Rusanen}, \citenamefont {V{\"a}h{\"a}kainu}, \citenamefont {Weckstr{\"o}m},\
  and\ \citenamefont {Arikawa}}]{rusanen2017characterization}%
  \BibitemOpen
  \bibfield  {author} {\bibinfo {author} {\bibfnamefont {J.}~\bibnamefont
  {Rusanen}}, \bibinfo {author} {\bibfnamefont {A.}~\bibnamefont
  {V{\"a}h{\"a}kainu}}, \bibinfo {author} {\bibfnamefont {M.}~\bibnamefont
  {Weckstr{\"o}m}}, \ and\ \bibinfo {author} {\bibfnamefont {K.}~\bibnamefont
  {Arikawa}},\ }\href@noop {} {\bibfield  {journal} {\bibinfo  {journal}
  {Journal of Comparative Physiology A}\ }\textbf {\bibinfo {volume} {203}},\
  \bibinfo {pages} {903} (\bibinfo {year} {2017})}\BibitemShut {NoStop}%
\bibitem [{\citenamefont {Mendelson}(1971)}]{mendel}%
  \BibitemOpen
  \bibfield  {author} {\bibinfo {author} {\bibfnamefont {M.}~\bibnamefont
  {Mendelson}},\ }\href@noop {} {\bibfield  {journal} {\bibinfo  {journal}
  {Science}\ }\textbf {\bibinfo {volume} {171}},\ \bibinfo {pages} {1170}
  (\bibinfo {year} {1971})}\BibitemShut {NoStop}%
\bibitem [{\citenamefont {DiCaprio}(2004)}]{dicaprio}%
  \BibitemOpen
  \bibfield  {author} {\bibinfo {author} {\bibfnamefont {R.~A.}\ \bibnamefont
  {DiCaprio}},\ }\href@noop {} {\bibfield  {journal} {\bibinfo  {journal}
  {Journal of neurophysiology}\ }\textbf {\bibinfo {volume} {92}},\ \bibinfo
  {pages} {302} (\bibinfo {year} {2004})}\BibitemShut {NoStop}%
\bibitem [{\citenamefont {Burrows}(1980)}]{burr}%
  \BibitemOpen
  \bibfield  {author} {\bibinfo {author} {\bibfnamefont {M.}~\bibnamefont
  {Burrows}},\ }\href@noop {} {\bibfield  {journal} {\bibinfo  {journal} {The
  Journal of physiology}\ }\textbf {\bibinfo {volume} {298}},\ \bibinfo {pages}
  {213} (\bibinfo {year} {1980})}\BibitemShut {NoStop}%
\bibitem [{\citenamefont {Beggs}\ and\ \citenamefont
  {Plenz}(2003)}]{beggsPlenz2003}%
  \BibitemOpen
  \bibfield  {author} {\bibinfo {author} {\bibfnamefont {J.~M.}\ \bibnamefont
  {Beggs}}\ and\ \bibinfo {author} {\bibfnamefont {D.}~\bibnamefont {Plenz}},\
  }\href@noop {} {\bibfield  {journal} {\bibinfo  {journal} {Journal of
  neuroscience}\ }\textbf {\bibinfo {volume} {23}},\ \bibinfo {pages} {11167}
  (\bibinfo {year} {2003})}\BibitemShut {NoStop}%
\bibitem [{\citenamefont {Mazzoni}\ \emph {et~al.}(2007)\citenamefont
  {Mazzoni}, \citenamefont {Broccard}, \citenamefont {Garcia-Perez},
  \citenamefont {Bonifazi}, \citenamefont {Ruaro},\ and\ \citenamefont
  {Torre}}]{mazzoni2007}%
  \BibitemOpen
  \bibfield  {author} {\bibinfo {author} {\bibfnamefont {A.}~\bibnamefont
  {Mazzoni}}, \bibinfo {author} {\bibfnamefont {F.~D.}\ \bibnamefont
  {Broccard}}, \bibinfo {author} {\bibfnamefont {E.}~\bibnamefont
  {Garcia-Perez}}, \bibinfo {author} {\bibfnamefont {P.}~\bibnamefont
  {Bonifazi}}, \bibinfo {author} {\bibfnamefont {M.~E.}\ \bibnamefont {Ruaro}},
  \ and\ \bibinfo {author} {\bibfnamefont {V.}~\bibnamefont {Torre}},\
  }\href@noop {} {\bibfield  {journal} {\bibinfo  {journal} {PloS one}\
  }\textbf {\bibinfo {volume} {2}},\ \bibinfo {pages} {e439} (\bibinfo {year}
  {2007})}\BibitemShut {NoStop}%
\bibitem [{\citenamefont {Pasquale}\ \emph {et~al.}(2008)\citenamefont
  {Pasquale}, \citenamefont {Massobrio}, \citenamefont {Bologna}, \citenamefont
  {Chiappalone},\ and\ \citenamefont {Martinoia}}]{pasquale2008}%
  \BibitemOpen
  \bibfield  {author} {\bibinfo {author} {\bibfnamefont {V.}~\bibnamefont
  {Pasquale}}, \bibinfo {author} {\bibfnamefont {P.}~\bibnamefont {Massobrio}},
  \bibinfo {author} {\bibfnamefont {L.}~\bibnamefont {Bologna}}, \bibinfo
  {author} {\bibfnamefont {M.}~\bibnamefont {Chiappalone}}, \ and\ \bibinfo
  {author} {\bibfnamefont {S.}~\bibnamefont {Martinoia}},\ }\href@noop {}
  {\bibfield  {journal} {\bibinfo  {journal} {Neuroscience}\ }\textbf {\bibinfo
  {volume} {153}},\ \bibinfo {pages} {1354} (\bibinfo {year}
  {2008})}\BibitemShut {NoStop}%
\bibitem [{\citenamefont {Gireesh}\ and\ \citenamefont
  {Plenz}(2008)}]{gireesh2008}%
  \BibitemOpen
  \bibfield  {author} {\bibinfo {author} {\bibfnamefont {E.~D.}\ \bibnamefont
  {Gireesh}}\ and\ \bibinfo {author} {\bibfnamefont {D.}~\bibnamefont
  {Plenz}},\ }\href@noop {} {\bibfield  {journal} {\bibinfo  {journal}
  {Proceedings of the National Academy of Sciences}\ }\textbf {\bibinfo
  {volume} {105}},\ \bibinfo {pages} {7576} (\bibinfo {year}
  {2008})}\BibitemShut {NoStop}%
\bibitem [{\citenamefont {Petermann}\ \emph {et~al.}(2009)\citenamefont
  {Petermann}, \citenamefont {Thiagarajan}, \citenamefont {Lebedev},
  \citenamefont {Nicolelis}, \citenamefont {Chialvo},\ and\ \citenamefont
  {Plenz}}]{petermann2009}%
  \BibitemOpen
  \bibfield  {author} {\bibinfo {author} {\bibfnamefont {T.}~\bibnamefont
  {Petermann}}, \bibinfo {author} {\bibfnamefont {T.~C.}\ \bibnamefont
  {Thiagarajan}}, \bibinfo {author} {\bibfnamefont {M.~A.}\ \bibnamefont
  {Lebedev}}, \bibinfo {author} {\bibfnamefont {M.~A.}\ \bibnamefont
  {Nicolelis}}, \bibinfo {author} {\bibfnamefont {D.~R.}\ \bibnamefont
  {Chialvo}}, \ and\ \bibinfo {author} {\bibfnamefont {D.}~\bibnamefont
  {Plenz}},\ }\href@noop {} {\bibfield  {journal} {\bibinfo  {journal}
  {Proceedings of the National Academy of Sciences}\ }\textbf {\bibinfo
  {volume} {106}},\ \bibinfo {pages} {15921} (\bibinfo {year}
  {2009})}\BibitemShut {NoStop}%
\bibitem [{\citenamefont {Haimovici}\ \emph {et~al.}(2013)\citenamefont
  {Haimovici}, \citenamefont {Tagliazucchi}, \citenamefont {Balenzuela},\ and\
  \citenamefont {Chialvo}}]{haimovici2013}%
  \BibitemOpen
  \bibfield  {author} {\bibinfo {author} {\bibfnamefont {A.}~\bibnamefont
  {Haimovici}}, \bibinfo {author} {\bibfnamefont {E.}~\bibnamefont
  {Tagliazucchi}}, \bibinfo {author} {\bibfnamefont {P.}~\bibnamefont
  {Balenzuela}}, \ and\ \bibinfo {author} {\bibfnamefont {D.~R.}\ \bibnamefont
  {Chialvo}},\ }\href@noop {} {\bibfield  {journal} {\bibinfo  {journal}
  {Physical review letters}\ }\textbf {\bibinfo {volume} {110}},\ \bibinfo
  {pages} {178101} (\bibinfo {year} {2013})}\BibitemShut {NoStop}%
\bibitem [{\citenamefont {Shriki}\ \emph
  {et~al.}(2013{\natexlab{a}})\citenamefont {Shriki}, \citenamefont {Alstott},
  \citenamefont {Carver}, \citenamefont {Holroyd}, \citenamefont {Henson},
  \citenamefont {Smith}, \citenamefont {Coppola}, \citenamefont {Bullmore},\
  and\ \citenamefont {Plenz}}]{shriki2013}%
  \BibitemOpen
  \bibfield  {author} {\bibinfo {author} {\bibfnamefont {O.}~\bibnamefont
  {Shriki}}, \bibinfo {author} {\bibfnamefont {J.}~\bibnamefont {Alstott}},
  \bibinfo {author} {\bibfnamefont {F.}~\bibnamefont {Carver}}, \bibinfo
  {author} {\bibfnamefont {T.}~\bibnamefont {Holroyd}}, \bibinfo {author}
  {\bibfnamefont {R.~N.}\ \bibnamefont {Henson}}, \bibinfo {author}
  {\bibfnamefont {M.~L.}\ \bibnamefont {Smith}}, \bibinfo {author}
  {\bibfnamefont {R.}~\bibnamefont {Coppola}}, \bibinfo {author} {\bibfnamefont
  {E.}~\bibnamefont {Bullmore}}, \ and\ \bibinfo {author} {\bibfnamefont
  {D.}~\bibnamefont {Plenz}},\ }\href@noop {} {\bibfield  {journal} {\bibinfo
  {journal} {Journal of Neuroscience}\ }\textbf {\bibinfo {volume} {33}},\
  \bibinfo {pages} {7079} (\bibinfo {year} {2013}{\natexlab{a}})}\BibitemShut
  {NoStop}%
\bibitem [{\citenamefont {Massobrio}\ \emph {et~al.}(2015)\citenamefont
  {Massobrio}, \citenamefont {de~Arcangelis}, \citenamefont {Pasquale},
  \citenamefont {Jensen},\ and\ \citenamefont {Plenz}}]{critrev}%
  \BibitemOpen
  \bibfield  {author} {\bibinfo {author} {\bibfnamefont {P.}~\bibnamefont
  {Massobrio}}, \bibinfo {author} {\bibfnamefont {L.}~\bibnamefont
  {de~Arcangelis}}, \bibinfo {author} {\bibfnamefont {V.}~\bibnamefont
  {Pasquale}}, \bibinfo {author} {\bibfnamefont {H.~J.}\ \bibnamefont
  {Jensen}}, \ and\ \bibinfo {author} {\bibfnamefont {D.}~\bibnamefont
  {Plenz}},\ }\href@noop {} {\bibfield  {journal} {\bibinfo  {journal}
  {Frontiers in Systems Neuroscience}\ }\textbf {\bibinfo {volume} {9}},\
  \bibinfo {pages} {1} (\bibinfo {year} {2015})}\BibitemShut {NoStop}%
\bibitem [{\citenamefont {Zapperi}\ \emph {et~al.}(1995)\citenamefont
  {Zapperi}, \citenamefont {Lauritsen},\ and\ \citenamefont
  {Stanley}}]{zapperi1995}%
  \BibitemOpen
  \bibfield  {author} {\bibinfo {author} {\bibfnamefont {S.}~\bibnamefont
  {Zapperi}}, \bibinfo {author} {\bibfnamefont {K.~B.}\ \bibnamefont
  {Lauritsen}}, \ and\ \bibinfo {author} {\bibfnamefont {H.~E.}\ \bibnamefont
  {Stanley}},\ }\href@noop {} {\bibfield  {journal} {\bibinfo  {journal}
  {Physical review letters}\ }\textbf {\bibinfo {volume} {75}},\ \bibinfo
  {pages} {4071} (\bibinfo {year} {1995})}\BibitemShut {NoStop}%
\bibitem [{\citenamefont {de~Arcangelis}\ \emph {et~al.}(2006)\citenamefont
  {de~Arcangelis}, \citenamefont {Perrone-Capano},\ and\ \citenamefont
  {Herrmann}}]{lucilla2006}%
  \BibitemOpen
  \bibfield  {author} {\bibinfo {author} {\bibfnamefont {L.}~\bibnamefont
  {de~Arcangelis}}, \bibinfo {author} {\bibfnamefont {C.}~\bibnamefont
  {Perrone-Capano}}, \ and\ \bibinfo {author} {\bibfnamefont {H.~J.}\
  \bibnamefont {Herrmann}},\ }\href@noop {} {\bibfield  {journal} {\bibinfo
  {journal} {Physical review letters}\ }\textbf {\bibinfo {volume} {96}},\
  \bibinfo {pages} {028107} (\bibinfo {year} {2006})}\BibitemShut {NoStop}%
\bibitem [{\citenamefont {Levina}\ \emph {et~al.}(2007)\citenamefont {Levina},
  \citenamefont {Herrmann},\ and\ \citenamefont {Geisel}}]{levina}%
  \BibitemOpen
  \bibfield  {author} {\bibinfo {author} {\bibfnamefont {A.}~\bibnamefont
  {Levina}}, \bibinfo {author} {\bibfnamefont {J.~M.}\ \bibnamefont
  {Herrmann}}, \ and\ \bibinfo {author} {\bibfnamefont {T.}~\bibnamefont
  {Geisel}},\ }\href@noop {} {\bibfield  {journal} {\bibinfo  {journal} {Nature
  physics}\ }\textbf {\bibinfo {volume} {3}},\ \bibinfo {pages} {857} (\bibinfo
  {year} {2007})}\BibitemShut {NoStop}%
\bibitem [{\citenamefont {Kanders}\ \emph {et~al.}(2017)\citenamefont
  {Kanders}, \citenamefont {Lorimer},\ and\ \citenamefont {Stoop}}]{kanders}%
  \BibitemOpen
  \bibfield  {author} {\bibinfo {author} {\bibfnamefont {K.}~\bibnamefont
  {Kanders}}, \bibinfo {author} {\bibfnamefont {T.}~\bibnamefont {Lorimer}}, \
  and\ \bibinfo {author} {\bibfnamefont {R.}~\bibnamefont {Stoop}},\
  }\href@noop {} {\bibfield  {journal} {\bibinfo  {journal} {Chaos: An
  Interdisciplinary Journal of Nonlinear Science}\ }\textbf {\bibinfo {volume}
  {27}},\ \bibinfo {pages} {047408} (\bibinfo {year} {2017})}\BibitemShut
  {NoStop}%
\bibitem [{\citenamefont {de~Arcangelis}(2012)}]{dragon}%
  \BibitemOpen
  \bibfield  {author} {\bibinfo {author} {\bibfnamefont {L.}~\bibnamefont
  {de~Arcangelis}},\ }\href@noop {} {\bibfield  {journal} {\bibinfo  {journal}
  {The European Physical Journal. Special Topics}\ }\textbf {\bibinfo {volume}
  {205}},\ \bibinfo {pages} {243} (\bibinfo {year} {2012})}\BibitemShut
  {NoStop}%
\bibitem [{\citenamefont {Lombardi}\ \emph {et~al.}(2017)\citenamefont
  {Lombardi}, \citenamefont {Herrmann},\ and\ \citenamefont
  {de~Arcangelis}}]{lombchaos}%
  \BibitemOpen
  \bibfield  {author} {\bibinfo {author} {\bibfnamefont {F.}~\bibnamefont
  {Lombardi}}, \bibinfo {author} {\bibfnamefont {H.}~\bibnamefont {Herrmann}},
  \ and\ \bibinfo {author} {\bibfnamefont {L.}~\bibnamefont {de~Arcangelis}},\
  }\href@noop {} {\bibfield  {journal} {\bibinfo  {journal} {Chaos: An
  Interdisciplinary Journal of Nonlinear Science}\ }\textbf {\bibinfo {volume}
  {27}},\ \bibinfo {pages} {1} (\bibinfo {year} {2017})}\BibitemShut {NoStop}%
\bibitem [{\citenamefont {de~Arcangelis}\ and\ \citenamefont
  {Herrmann}(2012)}]{lucillaActD}%
  \BibitemOpen
  \bibfield  {author} {\bibinfo {author} {\bibfnamefont {L.}~\bibnamefont
  {de~Arcangelis}}\ and\ \bibinfo {author} {\bibfnamefont {H.~J.}\ \bibnamefont
  {Herrmann}},\ }\href@noop {} {\bibfield  {journal} {\bibinfo  {journal}
  {Frontiers in physiology}\ }\textbf {\bibinfo {volume} {3}},\ \bibinfo
  {pages} {1} (\bibinfo {year} {2012})}\BibitemShut {NoStop}%
\bibitem [{\citenamefont {van Kessenich}\ \emph {et~al.}(2018)\citenamefont
  {van Kessenich}, \citenamefont {Lukovi{\'c}}, \citenamefont {de~Arcangelis},\
  and\ \citenamefont {Herrmann}}]{van2018}%
  \BibitemOpen
  \bibfield  {author} {\bibinfo {author} {\bibfnamefont {L.~M.}\ \bibnamefont
  {van Kessenich}}, \bibinfo {author} {\bibfnamefont {M.}~\bibnamefont
  {Lukovi{\'c}}}, \bibinfo {author} {\bibfnamefont {L.}~\bibnamefont
  {de~Arcangelis}}, \ and\ \bibinfo {author} {\bibfnamefont {H.~J.}\
  \bibnamefont {Herrmann}},\ }\href@noop {} {\bibfield  {journal} {\bibinfo
  {journal} {Physical Review E}\ }\textbf {\bibinfo {volume} {97}},\ \bibinfo
  {pages} {032312} (\bibinfo {year} {2018})}\BibitemShut {NoStop}%
\bibitem [{\citenamefont {de~Arcangelis}\ and\ \citenamefont
  {Herrmann}(2010)}]{lucillaLearning}%
  \BibitemOpen
  \bibfield  {author} {\bibinfo {author} {\bibfnamefont {L.}~\bibnamefont
  {de~Arcangelis}}\ and\ \bibinfo {author} {\bibfnamefont {H.~J.}\ \bibnamefont
  {Herrmann}},\ }\href@noop {} {\bibfield  {journal} {\bibinfo  {journal}
  {Proceedings of the National Academy of Sciences}\ }\textbf {\bibinfo
  {volume} {107}},\ \bibinfo {pages} {3977} (\bibinfo {year}
  {2010})}\BibitemShut {NoStop}%
\bibitem [{\citenamefont {Capano}\ \emph {et~al.}(2015)\citenamefont {Capano},
  \citenamefont {Herrmann},\ and\ \citenamefont
  {de~Arcangelis}}]{vittorioLearning}%
  \BibitemOpen
  \bibfield  {author} {\bibinfo {author} {\bibfnamefont {V.}~\bibnamefont
  {Capano}}, \bibinfo {author} {\bibfnamefont {H.~J.}\ \bibnamefont
  {Herrmann}}, \ and\ \bibinfo {author} {\bibfnamefont {L.}~\bibnamefont
  {de~Arcangelis}},\ }\href@noop {} {\bibfield  {journal} {\bibinfo  {journal}
  {Scientific reports}\ }\textbf {\bibinfo {volume} {5}},\ \bibinfo {pages}
  {9895} (\bibinfo {year} {2015})}\BibitemShut {NoStop}%
\bibitem [{\citenamefont {Sarracino}\ \emph {et~al.}(2020)\citenamefont
  {Sarracino}, \citenamefont {Arviv}, \citenamefont {Shriki},\ and\
  \citenamefont {de~Arcangelis}}]{fdt}%
  \BibitemOpen
  \bibfield  {author} {\bibinfo {author} {\bibfnamefont {A.}~\bibnamefont
  {Sarracino}}, \bibinfo {author} {\bibfnamefont {O.}~\bibnamefont {Arviv}},
  \bibinfo {author} {\bibfnamefont {O.}~\bibnamefont {Shriki}}, \ and\ \bibinfo
  {author} {\bibfnamefont {L.}~\bibnamefont {de~Arcangelis}},\ }\href@noop {}
  {\bibfield  {journal} {\bibinfo  {journal} {Physical Review Research}\
  }\textbf {\bibinfo {volume} {2}},\ \bibinfo {pages} {033355} (\bibinfo {year}
  {2020})}\BibitemShut {NoStop}%
\bibitem [{\citenamefont {Janssen}(1981)}]{janssen1981nonequilibrium}%
  \BibitemOpen
  \bibfield  {author} {\bibinfo {author} {\bibfnamefont {H.-K.}\ \bibnamefont
  {Janssen}},\ }\href@noop {} {\bibfield  {journal} {\bibinfo  {journal}
  {Zeitschrift f{\"u}r Physik B Condensed Matter}\ }\textbf {\bibinfo {volume}
  {42}},\ \bibinfo {pages} {151} (\bibinfo {year} {1981})}\BibitemShut
  {NoStop}%
\bibitem [{\citenamefont {Grassberger}(1981)}]{grassberger1981phase}%
  \BibitemOpen
  \bibfield  {author} {\bibinfo {author} {\bibfnamefont {P.}~\bibnamefont
  {Grassberger}},\ }in\ \href@noop {} {\emph {\bibinfo {booktitle} {Nonlinear
  Phenomena in Chemical Dynamics}}}\ (\bibinfo  {publisher} {Springer},\
  \bibinfo {year} {1981})\ pp.\ \bibinfo {pages} {262--262}\BibitemShut
  {NoStop}%
\bibitem [{\citenamefont {Poil}\ \emph {et~al.}(2012)\citenamefont {Poil},
  \citenamefont {Hardstone}, \citenamefont {Mansvelder},\ and\ \citenamefont
  {Linkenkaer-Hansen}}]{poil2012critical}%
  \BibitemOpen
  \bibfield  {author} {\bibinfo {author} {\bibfnamefont {S.-S.}\ \bibnamefont
  {Poil}}, \bibinfo {author} {\bibfnamefont {R.}~\bibnamefont {Hardstone}},
  \bibinfo {author} {\bibfnamefont {H.~D.}\ \bibnamefont {Mansvelder}}, \ and\
  \bibinfo {author} {\bibfnamefont {K.}~\bibnamefont {Linkenkaer-Hansen}},\
  }\href@noop {} {\bibfield  {journal} {\bibinfo  {journal} {Journal of
  Neuroscience}\ }\textbf {\bibinfo {volume} {32}},\ \bibinfo {pages} {9817}
  (\bibinfo {year} {2012})}\BibitemShut {NoStop}%
\bibitem [{\citenamefont {Shriki}\ \emph
  {et~al.}(2013{\natexlab{b}})\citenamefont {Shriki}, \citenamefont {Alstott},
  \citenamefont {Carver}, \citenamefont {Holroyd}, \citenamefont {Henson},
  \citenamefont {Smith}, \citenamefont {Coppola}, \citenamefont {Bullmore},\
  and\ \citenamefont {Plenz}}]{shriki2013neuronal}%
  \BibitemOpen
  \bibfield  {author} {\bibinfo {author} {\bibfnamefont {O.}~\bibnamefont
  {Shriki}}, \bibinfo {author} {\bibfnamefont {J.}~\bibnamefont {Alstott}},
  \bibinfo {author} {\bibfnamefont {F.}~\bibnamefont {Carver}}, \bibinfo
  {author} {\bibfnamefont {T.}~\bibnamefont {Holroyd}}, \bibinfo {author}
  {\bibfnamefont {R.~N.}\ \bibnamefont {Henson}}, \bibinfo {author}
  {\bibfnamefont {M.~L.}\ \bibnamefont {Smith}}, \bibinfo {author}
  {\bibfnamefont {R.}~\bibnamefont {Coppola}}, \bibinfo {author} {\bibfnamefont
  {E.}~\bibnamefont {Bullmore}}, \ and\ \bibinfo {author} {\bibfnamefont
  {D.}~\bibnamefont {Plenz}},\ }\href@noop {} {\bibfield  {journal} {\bibinfo
  {journal} {Journal of Neuroscience}\ }\textbf {\bibinfo {volume} {33}},\
  \bibinfo {pages} {7079} (\bibinfo {year} {2013}{\natexlab{b}})}\BibitemShut
  {NoStop}%
\bibitem [{\citenamefont {Dalla~Porta}\ and\ \citenamefont
  {Copelli}(2019)}]{dalla2019modeling}%
  \BibitemOpen
  \bibfield  {author} {\bibinfo {author} {\bibfnamefont {L.}~\bibnamefont
  {Dalla~Porta}}\ and\ \bibinfo {author} {\bibfnamefont {M.}~\bibnamefont
  {Copelli}},\ }\href@noop {} {\bibfield  {journal} {\bibinfo  {journal} {PLoS
  computational biology}\ }\textbf {\bibinfo {volume} {15}},\ \bibinfo {pages}
  {e1006924} (\bibinfo {year} {2019})}\BibitemShut {NoStop}%
\bibitem [{\citenamefont {Palva}\ \emph {et~al.}(2013)\citenamefont {Palva},
  \citenamefont {Zhigalov}, \citenamefont {Hirvonen}, \citenamefont {Korhonen},
  \citenamefont {Linkenkaer-Hansen},\ and\ \citenamefont
  {Palva}}]{palva2013neuronal}%
  \BibitemOpen
  \bibfield  {author} {\bibinfo {author} {\bibfnamefont {J.~M.}\ \bibnamefont
  {Palva}}, \bibinfo {author} {\bibfnamefont {A.}~\bibnamefont {Zhigalov}},
  \bibinfo {author} {\bibfnamefont {J.}~\bibnamefont {Hirvonen}}, \bibinfo
  {author} {\bibfnamefont {O.}~\bibnamefont {Korhonen}}, \bibinfo {author}
  {\bibfnamefont {K.}~\bibnamefont {Linkenkaer-Hansen}}, \ and\ \bibinfo
  {author} {\bibfnamefont {S.}~\bibnamefont {Palva}},\ }\href@noop {}
  {\bibfield  {journal} {\bibinfo  {journal} {Proceedings of the National
  Academy of Sciences}\ }\textbf {\bibinfo {volume} {110}},\ \bibinfo {pages}
  {3585} (\bibinfo {year} {2013})}\BibitemShut {NoStop}%
\bibitem [{\citenamefont {Zhigalov}\ \emph {et~al.}(2015)\citenamefont
  {Zhigalov}, \citenamefont {Arnulfo}, \citenamefont {Nobili}, \citenamefont
  {Palva},\ and\ \citenamefont {Palva}}]{zhigalov2015relationship}%
  \BibitemOpen
  \bibfield  {author} {\bibinfo {author} {\bibfnamefont {A.}~\bibnamefont
  {Zhigalov}}, \bibinfo {author} {\bibfnamefont {G.}~\bibnamefont {Arnulfo}},
  \bibinfo {author} {\bibfnamefont {L.}~\bibnamefont {Nobili}}, \bibinfo
  {author} {\bibfnamefont {S.}~\bibnamefont {Palva}}, \ and\ \bibinfo {author}
  {\bibfnamefont {J.~M.}\ \bibnamefont {Palva}},\ }\href@noop {} {\bibfield
  {journal} {\bibinfo  {journal} {Journal of Neuroscience}\ }\textbf {\bibinfo
  {volume} {35}},\ \bibinfo {pages} {5385} (\bibinfo {year}
  {2015})}\BibitemShut {NoStop}%
\bibitem [{\citenamefont {Yaghoubi}\ \emph {et~al.}(2018)\citenamefont
  {Yaghoubi}, \citenamefont {de~Graaf}, \citenamefont {Orlandi}, \citenamefont
  {Girotto}, \citenamefont {Colicos},\ and\ \citenamefont
  {Davidsen}}]{yaghoubi2018neuronal}%
  \BibitemOpen
  \bibfield  {author} {\bibinfo {author} {\bibfnamefont {M.}~\bibnamefont
  {Yaghoubi}}, \bibinfo {author} {\bibfnamefont {T.}~\bibnamefont {de~Graaf}},
  \bibinfo {author} {\bibfnamefont {J.~G.}\ \bibnamefont {Orlandi}}, \bibinfo
  {author} {\bibfnamefont {F.}~\bibnamefont {Girotto}}, \bibinfo {author}
  {\bibfnamefont {M.~A.}\ \bibnamefont {Colicos}}, \ and\ \bibinfo {author}
  {\bibfnamefont {J.}~\bibnamefont {Davidsen}},\ }\href@noop {} {\bibfield
  {journal} {\bibinfo  {journal} {Scientific reports}\ }\textbf {\bibinfo
  {volume} {8}},\ \bibinfo {pages} {1} (\bibinfo {year} {2018})}\BibitemShut
  {NoStop}%
\bibitem [{\citenamefont {Zetterberg}\ \emph {et~al.}(1978)\citenamefont
  {Zetterberg}, \citenamefont {Kristiansson},\ and\ \citenamefont
  {Mossberg}}]{zetterberg1978performance}%
  \BibitemOpen
  \bibfield  {author} {\bibinfo {author} {\bibfnamefont {L.}~\bibnamefont
  {Zetterberg}}, \bibinfo {author} {\bibfnamefont {L.}~\bibnamefont
  {Kristiansson}}, \ and\ \bibinfo {author} {\bibfnamefont {K.}~\bibnamefont
  {Mossberg}},\ }\href@noop {} {\bibfield  {journal} {\bibinfo  {journal}
  {Biological cybernetics}\ }\textbf {\bibinfo {volume} {31}},\ \bibinfo
  {pages} {15} (\bibinfo {year} {1978})}\BibitemShut {NoStop}%
\bibitem [{\citenamefont {Kretzberg}\ \emph {et~al.}(2001)\citenamefont
  {Kretzberg}, \citenamefont {Warzecha},\ and\ \citenamefont
  {Egelhaaf}}]{kretzberg2001neural}%
  \BibitemOpen
  \bibfield  {author} {\bibinfo {author} {\bibfnamefont {J.}~\bibnamefont
  {Kretzberg}}, \bibinfo {author} {\bibfnamefont {A.-K.}\ \bibnamefont
  {Warzecha}}, \ and\ \bibinfo {author} {\bibfnamefont {M.}~\bibnamefont
  {Egelhaaf}},\ }\href@noop {} {\bibfield  {journal} {\bibinfo  {journal}
  {Journal of computational neuroscience}\ }\textbf {\bibinfo {volume} {11}},\
  \bibinfo {pages} {153} (\bibinfo {year} {2001})}\BibitemShut {NoStop}%
\bibitem [{\citenamefont {Moosavi}\ \emph {et~al.}(2017)\citenamefont
  {Moosavi}, \citenamefont {Montakhab},\ and\ \citenamefont
  {Valizadeh}}]{moosavi2017refractory}%
  \BibitemOpen
  \bibfield  {author} {\bibinfo {author} {\bibfnamefont {S.~A.}\ \bibnamefont
  {Moosavi}}, \bibinfo {author} {\bibfnamefont {A.}~\bibnamefont {Montakhab}},
  \ and\ \bibinfo {author} {\bibfnamefont {A.}~\bibnamefont {Valizadeh}},\
  }\href@noop {} {\bibfield  {journal} {\bibinfo  {journal} {Scientific
  reports}\ }\textbf {\bibinfo {volume} {7}},\ \bibinfo {pages} {7107}
  (\bibinfo {year} {2017})}\BibitemShut {NoStop}%
\bibitem [{\citenamefont {Restrepo}\ \emph {et~al.}(2007)\citenamefont
  {Restrepo}, \citenamefont {Ott},\ and\ \citenamefont
  {Hunt}}]{restrepo2007approximating}%
  \BibitemOpen
  \bibfield  {author} {\bibinfo {author} {\bibfnamefont {J.~G.}\ \bibnamefont
  {Restrepo}}, \bibinfo {author} {\bibfnamefont {E.}~\bibnamefont {Ott}}, \
  and\ \bibinfo {author} {\bibfnamefont {B.~R.}\ \bibnamefont {Hunt}},\
  }\href@noop {} {\bibfield  {journal} {\bibinfo  {journal} {Physical Review
  E}\ }\textbf {\bibinfo {volume} {76}},\ \bibinfo {pages} {056119} (\bibinfo
  {year} {2007})}\BibitemShut {NoStop}%
\bibitem [{\citenamefont {Larremore}\ \emph {et~al.}(2012)\citenamefont
  {Larremore}, \citenamefont {Carpenter}, \citenamefont {Ott},\ and\
  \citenamefont {Restrepo}}]{larremore2012statistical}%
  \BibitemOpen
  \bibfield  {author} {\bibinfo {author} {\bibfnamefont {D.~B.}\ \bibnamefont
  {Larremore}}, \bibinfo {author} {\bibfnamefont {M.~Y.}\ \bibnamefont
  {Carpenter}}, \bibinfo {author} {\bibfnamefont {E.}~\bibnamefont {Ott}}, \
  and\ \bibinfo {author} {\bibfnamefont {J.~G.}\ \bibnamefont {Restrepo}},\
  }\href@noop {} {\bibfield  {journal} {\bibinfo  {journal} {Physical Review
  E}\ }\textbf {\bibinfo {volume} {85}},\ \bibinfo {pages} {066131} (\bibinfo
  {year} {2012})}\BibitemShut {NoStop}%
\bibitem [{\citenamefont {Alstr{\o}m}(1988)}]{alstrom1988mean}%
  \BibitemOpen
  \bibfield  {author} {\bibinfo {author} {\bibfnamefont {P.}~\bibnamefont
  {Alstr{\o}m}},\ }\href@noop {} {\bibfield  {journal} {\bibinfo  {journal}
  {Physical Review A}\ }\textbf {\bibinfo {volume} {38}},\ \bibinfo {pages}
  {4905} (\bibinfo {year} {1988})}\BibitemShut {NoStop}%
\bibitem [{\citenamefont {Najafi}\ and\ \citenamefont
  {Rahimi-Majd}(2019)}]{najafi2019effect}%
  \BibitemOpen
  \bibfield  {author} {\bibinfo {author} {\bibfnamefont {M.}~\bibnamefont
  {Najafi}}\ and\ \bibinfo {author} {\bibfnamefont {M.}~\bibnamefont
  {Rahimi-Majd}},\ }\href@noop {} {\bibfield  {journal} {\bibinfo  {journal}
  {Physica Scripta}\ }\textbf {\bibinfo {volume} {94}},\ \bibinfo {pages}
  {055208} (\bibinfo {year} {2019})}\BibitemShut {NoStop}%
\bibitem [{\citenamefont {Sethna}\ \emph {et~al.}(2001)\citenamefont {Sethna},
  \citenamefont {Dahmen},\ and\ \citenamefont {Myers}}]{sethna}%
  \BibitemOpen
  \bibfield  {author} {\bibinfo {author} {\bibfnamefont {J.~P.}\ \bibnamefont
  {Sethna}}, \bibinfo {author} {\bibfnamefont {K.~A.}\ \bibnamefont {Dahmen}},
  \ and\ \bibinfo {author} {\bibfnamefont {C.~R.}\ \bibnamefont {Myers}},\
  }\href@noop {} {\bibfield  {journal} {\bibinfo  {journal} {Nature}\ }\textbf
  {\bibinfo {volume} {410}},\ \bibinfo {pages} {242} (\bibinfo {year}
  {2001})}\BibitemShut {NoStop}%
\bibitem [{\citenamefont {Pass}\ \emph {et~al.}(2020)\citenamefont {Pass},
  \citenamefont {Assfalg}, \citenamefont {Tolve}, \citenamefont {Blaess},
  \citenamefont {Rothermel}, \citenamefont {Wiesner},\ and\ \citenamefont
  {Ricke}}]{pass}%
  \BibitemOpen
  \bibfield  {author} {\bibinfo {author} {\bibfnamefont {T.}~\bibnamefont
  {Pass}}, \bibinfo {author} {\bibfnamefont {M.}~\bibnamefont {Assfalg}},
  \bibinfo {author} {\bibfnamefont {M.}~\bibnamefont {Tolve}}, \bibinfo
  {author} {\bibfnamefont {S.}~\bibnamefont {Blaess}}, \bibinfo {author}
  {\bibfnamefont {M.}~\bibnamefont {Rothermel}}, \bibinfo {author}
  {\bibfnamefont {R.~J.}\ \bibnamefont {Wiesner}}, \ and\ \bibinfo {author}
  {\bibfnamefont {K.~M.}\ \bibnamefont {Ricke}},\ }\href@noop {} {\bibfield
  {journal} {\bibinfo  {journal} {Mol Neurobiol.}\ }\textbf {\bibinfo {volume}
  {57}},\ \bibinfo {pages} {3646–3657} (\bibinfo {year} {2020})}\BibitemShut
  {NoStop}%
\end{thebibliography}%

\end{document}